\documentclass[sigplan,nonacm]{acmart}

\usepackage{graphicx}
\usepackage{xspace}
\usepackage{amsmath}
\usepackage{subcaption}
\usepackage{booktabs}
\usepackage{multirow}
\usepackage{makecell}
\usepackage{pifont}
\usepackage[dvipsnames,table]{xcolor}
\usepackage{enumitem}
\usepackage{url}
\usepackage{xurl}  

\newcommand{\sys}{Coral\xspace}
\newcommand{\smallscale}{core setup\xspace}
\newcommand{\largescale}{extended setup\xspace}
\newcommand{\Smallscale}{Core setup\xspace}
\newcommand{\Largescale}{Extended setup\xspace}
\newcommand{\homo}{Homo\xspace}

\newcommand{\myparagraph}[1]{\par\addvspace{0.5em}\noindent\textbf{#1}}

\newcommand{\yes}{\textcolor{ForestGreen}{\ding{51}}}
\newcommand{\no}{\textcolor{BrickRed}{\ding{55}}}

\newcommand{\circledone}{\ding{172}}
\newcommand{\circledtwo}{\ding{173}}

\begin{document}





\title{\sys: Cost-Efficient Multi-LLM Serving over Heterogeneous Cloud GPUs}

\author{Yixuan Mei}
\email{yixuanm@andrew.cmu.edu}
\affiliation{
    \institution{Carnegie Mellon University}
    \city{Pittsburgh}
    \state{PA}
    \country{USA}
}

\author{Zikun Li}
\email{zikunl@andrew.cmu.edu}
\affiliation{
    \institution{Carnegie Mellon University}
    \city{Pittsburgh}
    \state{PA}
    \country{USA}
}

\author{Zixuan Chen}
\email{zixuanc3@andrew.cmu.edu}
\affiliation{
    \institution{Carnegie Mellon University}
    \city{Pittsburgh}
    \state{PA}
    \country{USA}
}

\author{Shiqi Pan}
\email{shiqip@andrew.cmu.edu}
\affiliation{
    \institution{Carnegie Mellon University}
    \city{Pittsburgh}
    \state{PA}
    \country{USA}
}

\author{Mengdi Wu}
\email{mengdiwu@andrew.cmu.edu}
\affiliation{
    \institution{Carnegie Mellon University}
    \city{Pittsburgh}
    \state{PA}
    \country{USA}
}

\author{Xupeng Miao}
\authornote{Work done at Purdue.}
\email{xupeng.miao@pku.edu.cn}
\affiliation{
    \institution{Peking University}
    \city{Beijing}
    \country{China}
}

\author{Zhihao Jia}
\email{zhihao@cmu.edu}
\affiliation{
    \institution{Carnegie Mellon University}
    \city{Pittsburgh}
    \state{PA}
    \country{USA}
}

\author{K. V. Rashmi}
\email{rvinayak@andrew.cmu.edu}
\affiliation{
    \institution{Carnegie Mellon University}
    \city{Pittsburgh}
    \state{PA}
    \country{USA}
}

\begin{abstract}
The landscape of large language models (LLMs) has become increasingly fragmented, with no single model dominating the field. Meanwhile, cloud providers offer a diverse array of mid-tier and older-generation GPUs. These heterogeneous resources provide better availability and deliver comparable performance-per-dollar to top-tier hardware. To efficiently harness these GPUs for concurrent multi-LLM serving, we introduce \sys: an adaptive, heterogeneity-aware serving system.
The key idea behind \sys is the joint optimization of resource allocation and replica serving strategies across all models. To keep pace with shifting throughput demands and resource availability, \sys employs a lossless, two-stage decomposition. This approach preserves joint optimality while reducing online solve times from hours to mere tens of seconds. Evaluations across 6 models and 20 GPU configurations demonstrate that \sys reduces serving costs by up to 2.79$\times$ compared to the strongest baseline, while delivering up to 2.39$\times$ higher goodput under scarce resource constraints.
\end{abstract}

\maketitle

\section{Introduction}
\label{sec:intro}


Large language models (LLMs) now power an expanding array of tasks, including interactive chatbots~\cite{openai_chatgpt, google_gemini}, automated code generation~\cite{chen2021evaluating, zhu2024deepseek, roziere2023code}, and agentic workflows~\cite{yao2022react, xi2025rise}. Yet, no single model dominates the ecosystem. Models from different families excel at distinct tasks~\cite{liang2022holistic, chiang2024chatbot, jimenez2024swebench}, and models of varying sizes within those families target different cost--quality trade-offs~\cite{yang2025qwen3, jiang2024mixtral, meta2025llama4}. Industry data underscores this fragmentation: no single model handles more than 25\% of queries~\cite{perplexity2026modelswitching}. As a result, providers must efficiently serve multiple models concurrently. Perplexity, for example, serves 46 models for its answer engine~\cite{perplexity2026modelswitching}, and Microsoft Office 365 backs its AI features with seven LLMs across three families~\cite{jaiswal2025sageserve}.

To avoid the prohibitive capital costs of dedicated AI hardware, organizations increasingly rely on public clouds to host this diverse mix of models~\cite{openai2023azure, anthropic2023aws}. These cloud environments, however, are inherently heterogeneous, offering GPUs that span multiple generations and specifications (Table~\ref{tab:gpu-specs}). While top-tier GPUs (e.g., B200, H100) deliver peak per-device capability, they are often supply-constrained. In contrast, older and mid-tier GPUs (e.g., A100, L4) are more readily available, substantially cheaper, and often yield comparable or even superior cost efficiency per unit of compute and memory.


\begin{table}
\centering
\footnotesize
\setlength{\tabcolsep}{3pt}
\renewcommand{\arraystretch}{1.15}
\begin{tabular}{@{}l|c|c|c|c|c|c|c|c|c|c@{}}
\toprule
\multirow{2}{*}{\textbf{GPU}}
  & \multirow{2}{*}{\textbf{A}}
  & \multirow{2}{*}{\textbf{G}}
  & \multirow{2}{*}{\textbf{R}}
  & \multirow{2}{*}{\makecell{\textbf{Rel.}\\\textbf{Cost}}}
  & \multirow{2}{*}{\makecell{\textbf{Mem}\\\textbf{(GB)}}}
  & \multirow{2}{*}{\makecell{\textbf{BW}\\\textbf{(TB/s)}}}
  & \multirow{2}{*}{\makecell{\textbf{TF}\\\textbf{LOPS}}}
  & \multicolumn{3}{c@{}}{\textbf{Perf. Per Cost}} \\
  \cline{9-11}
  & & & & & & & & \textbf{Mem} & \textbf{BW} & \textbf{TF} \\
\midrule
H100 & \yes & \yes & \yes & 7.6 &  80 & 3.35 &  989 & 10.5 & 0.44 & \cellcolor[HTML]{DCEBDC}129.8 \\
A100 & \yes & \yes & \yes & 3.5 &  80 & 2.04 &  312 & \cellcolor[HTML]{DCEBDC}22.8 & \cellcolor[HTML]{A8D5A8}0.58 & 88.9 \\
L40S & \yes & \no  & \yes & 2.2 &  48 & 0.86 &  362 & 21.5 & 0.39 & \cellcolor[HTML]{A8D5A8}162.3 \\
L4   & \yes & \yes & \yes & 1.0 &  24 & 0.30 &  121 & \cellcolor[HTML]{A8D5A8}24.0 & 0.30 & 121.0 \\
A10G & \yes & \no  & \no  & 1.2 &  24 & 0.60 &   70 & 19.7 & \cellcolor[HTML]{DCEBDC}0.49 & 57.4 \\
\bottomrule
\end{tabular}
\caption{GPU specs and availabilities on AWS (A), GCP (G), and RunPod (R). Relative cost is mean hourly price normalized to L4. ``Perf. Per Cost'' divides metrics by relative cost.}
\label{tab:gpu-specs}
\vspace{-2em}
\end{table}

Together, model fragmentation and cloud GPU heterogeneity motivate a new challenge: \emph{cost-efficient, adaptive multi-LLM serving on heterogeneous cloud resources}. To address this, a serving system must take as input a set of models with latency service-level objectives (SLOs), current per-model throughput demands, and the real-time pricing and availability of GPU configurations across cloud regions. From these inputs, the system must jointly optimize two decisions: a \emph{resource allocation} plan detailing which nodes to provision, and a \emph{model placement} strategy dictating how to partition each model across those heterogeneous nodes. The objective is to satisfy all throughput demands and latency SLOs at the minimum total cost. Furthermore, because demand, resource availability, and pricing continuously fluctuate, the system must periodically re-solve this optimization to dynamically \emph{adapt} the cluster to shifting conditions.

Prior work tackles this problem only in pieces. Existing heterogeneity-aware allocation systems target orthogonal goals, such as spot preemption resilience via cross-region replica spreading~\cite{mao2025skyserve} or managing mixed interactive/batch workloads via forecast-driven scaling~\cite{jaiswal2025sageserve}. They share a common limitation: each model replica is treated as a black box on homogeneous hardware, abstracting model placement out of the optimization space entirely. Cauchy~\cite{zhang2025cauchy} relaxes this for prefill-decode disaggregated serving by assigning different GPU configurations to prefill and decode, but each phase remains internally homogeneous. As shown in Sec.~\ref{sec:opportunities} and Sec.~\ref{sec:evaluation}, this results not only in significantly higher costs but also in deployment infeasibility when resources are scarce.

Helix~\cite{mei2025helix} and HexGen~\cite{jiang2023hexgen} take the opposite stance, optimizing placement strictly for a \emph{single} model on a \emph{given} heterogeneous node set. Wrapping these single-model solvers in an outer enumeration over allocations is intractable. The allocation space is exponential, and each inner placement solve is already prohibitively expensive; Helix, for instance, takes four hours for one model on 24 nodes. As Table~\ref{tab:baselines} summarizes, no existing system addresses the problem end-to-end.

\begin{table}[t]
\centering
\footnotesize
\setlength{\tabcolsep}{4pt}
\renewcommand{\arraystretch}{1.15}
\begin{tabular}{@{}l|c|c|c|c|c|c@{}}
\toprule
\textbf{Criterion} & \textbf{\sys}
  & \makecell{\textbf{SkyS.}\\\cite{mao2025skyserve}}
  & \makecell{\textbf{Sage.}\\\cite{jaiswal2025sageserve}}
  & \makecell{\textbf{Cauchy}\\\cite{zhang2025cauchy}}
  & \makecell{\textbf{Helix}\\\cite{mei2025helix}}
  & \makecell{\textbf{HexG.}\\\cite{jiang2023hexgen}} \\
\midrule
Resource Alloc.  & \multirow{2}{*}{\makecell{Joint\\Opt.}} & \yes & \yes & \yes & \no  & \no  \\
Model Placement  &                                          & \no  & \no  & \no  & \yes & \yes \\
\midrule
Latency SLO      & \yes & \no  & \yes & \yes & \no  & \yes \\
Multi-LLM        & \yes & \no  & \yes & \yes  & \no  & \no  \\
\bottomrule
\end{tabular}
\caption{Comparison of heterogeneous LLM serving systems. Prior work optimizes either resource allocation~\cite{mao2025skyserve, jaiswal2025sageserve, zhang2025cauchy} or model placement~\cite{mei2025helix, jiang2023hexgen}. \sys uniquely co-optimizes both to serve multiple LLMs under per-model latency SLOs.}
\label{tab:baselines}
\vspace{-2em}
\end{table}

Solving the heterogeneity-aware multi-LLM serving problem end-to-end poses two key challenges. \emph{First, resource allocation and model placement are tightly coupled, and each features an exponential solution space.} The throughput and SLO attainable on a given set of nodes depend entirely on how we place the model. Conversely, the optimal placement strategy cannot be determined until the node allocation is finalized. Neither dimension can be pruned independently without prematurely constraining the other, yet a joint brute-force approach remains intractable due to the combinatorially explosive search space. \emph{Second, optimization must complete in minutes, not hours.} Because cloud GPU availability~\cite{wu2024can, strati2025sailor} and user throughput demands~\cite{stojkovic2025dynamollm} fluctuate rapidly, a computationally expensive solution that takes hours to generate will be rendered obsolete by the time it is deployed.

To address these challenges, we introduce \sys\footnote{\sys stands for \textbf{C}ost-efficient \textbf{O}rchestration of \textbf{R}esources for \textbf{A}daptive \textbf{L}LM-serving.}: an adaptive, heterogeneity-aware multi-LLM serving system. \sys is grounded in a key observation: for a given model and latency SLO, the throughput-optimal placement on any combination of nodes depends strictly on that combination, independent of the broader cluster allocation. Consequently, this optimal placement can be treated as a reusable artifact—which we term a \emph{Serving Template}—that can be \emph{pre-computed offline and cached}. This insight decouples model placement from resource allocation without sacrificing joint optimality.

\begin{figure}
    \begin{subfigure}{0.48\linewidth}
    \centering
    \includegraphics[width=\linewidth]{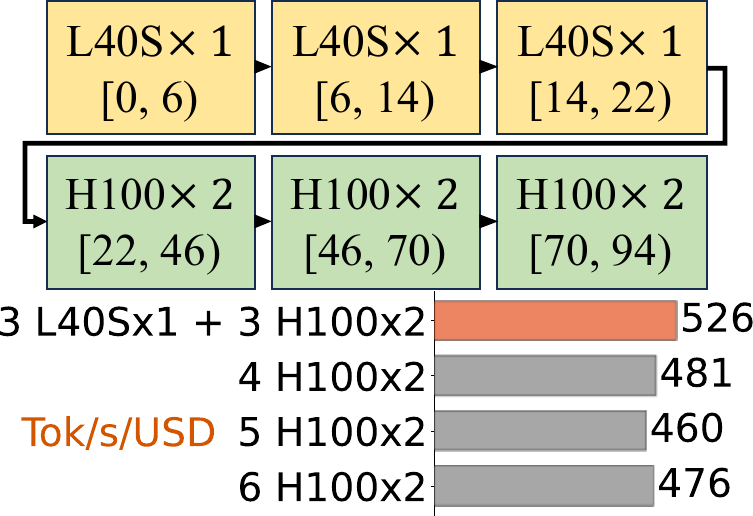}
    \caption{A mixed L40S/H100 pipeline serves Qwen-3 235B prefill more cost-efficiently than any pure-H100 setup (SLO = 1800 ms).}
    \label{fig:sec2_challenges_mixed_slo}
    \end{subfigure}
    \hfill
    \begin{subfigure}{0.48\linewidth}
    \centering
    \includegraphics[width=\linewidth]{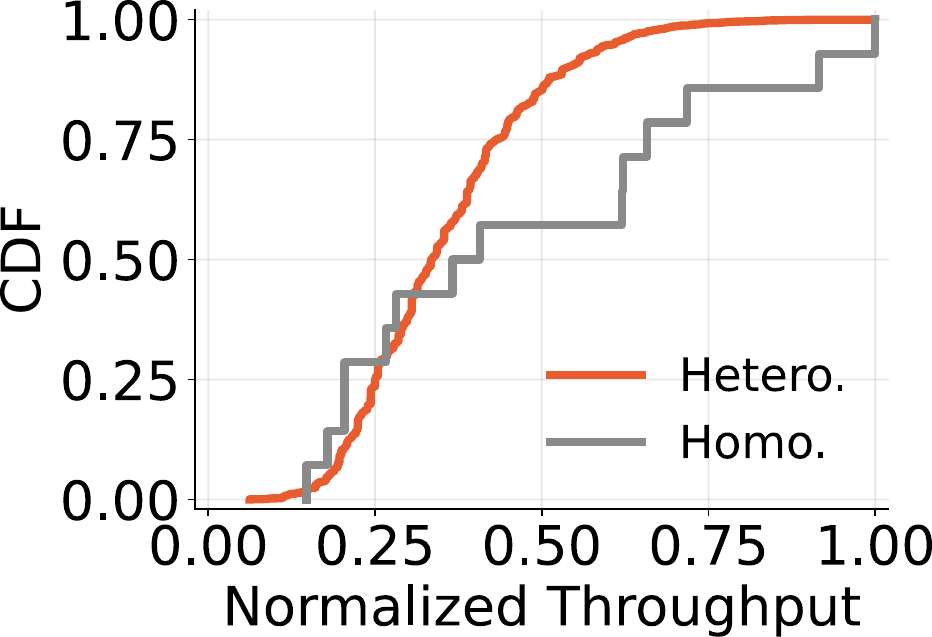}
    \caption{Normalized throughput CDF for GPT-OSS 120B decode plans. Heterogeneous combinations fill gaps left by homogeneous ones.}
    \label{fig:sec2_challenges_mixed_throughput}
    \end{subfigure}
    \caption{Opportunities brought by heterogeneity.}
    \label{fig:sec2_challenges_heterogeneity}
    \Description{Opportunities brought by heterogeneity.}
\end{figure}
\begin{figure}
    \centering
    \includegraphics[width=\linewidth]{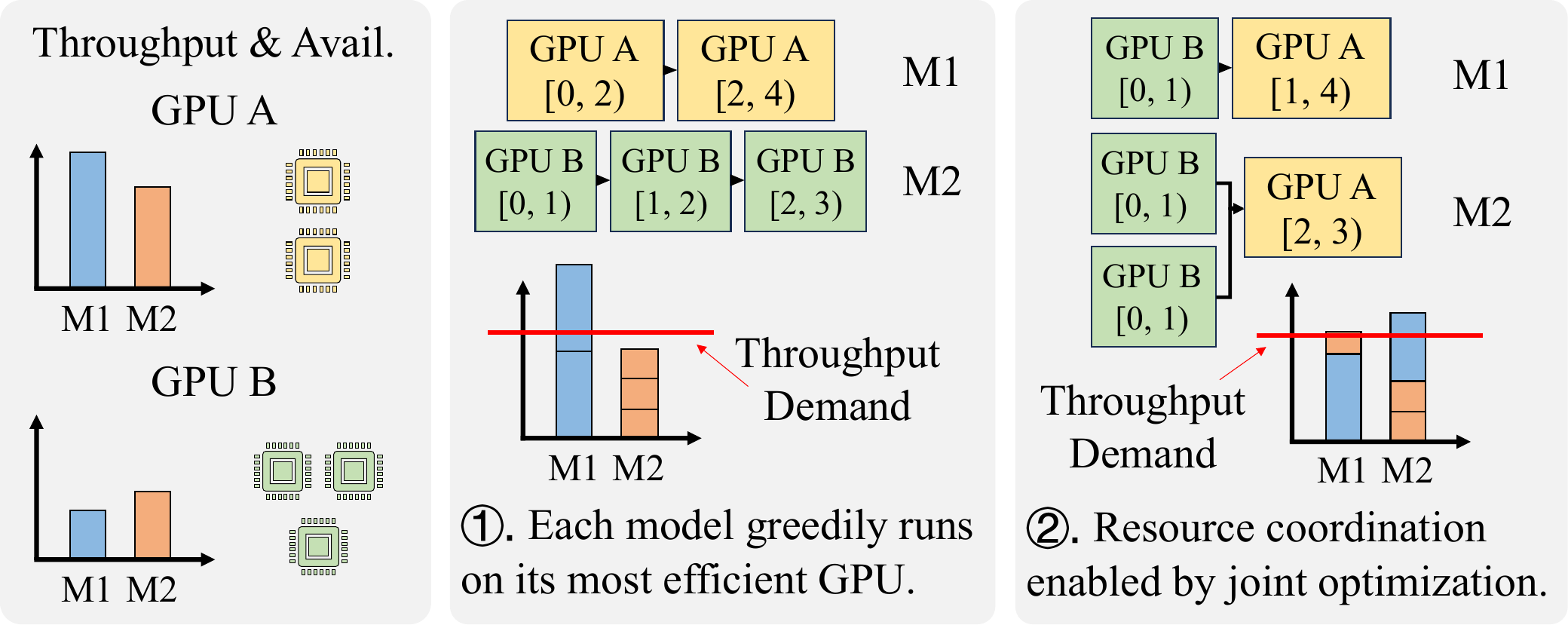}
    \caption{Joint optimization across models. Greedy per-model allocation (\circledone) causes contention; mixed-GPU replicas produced by joint optimization (\circledtwo) satisfy both demands.}
    \label{fig:sec2_challenges_multi_llm}
    \Description{Joint optimization across models.}
\end{figure}

While the theoretical space of node combinations is unbounded, \sys enumerates a principled subset that, as demonstrated in Sec.~\ref{sec:sensitivity}, captures the cost-efficient regime with negligible loss. Accordingly, \sys adopts a two-stage architecture. Offline, a template generator explores node combinations for each (model, SLO) pair, leveraging an Integer Linear Program (ILP) to compute the throughput-optimal placement under SLO and produce a library of reusable Serving Templates. Online, a lightweight ILP dynamically deploys these templates across cloud regions to satisfy throughput demands at minimum cost, factoring in real-time pricing and availability. By lifting the computationally heavy placement search off the critical path, the online solver completes in tens of seconds, ensuring the cluster remains highly adaptive to shifting demands and resource fluctuations.

We implement \sys in 53K lines of Python and C++, building a distributed runtime atop ZeroMQ~\cite{hintjens2013zeromq} and NCCL~\cite{nvidia_nccl}, with vLLM~\cite{kwon2023efficient} for per-node execution. To evaluate large-scale deployments without prohibitive hardware costs, we also develop a high-fidelity, event-driven simulator. Evaluated across 6 models, 5 GPU types, 20 GPU configurations, and varying throughput demands, \sys reduces total serving costs by up to 2.79$\times$ over the strongest baseline. Under strict resource constraints, it delivers up to 2.39$\times$ higher goodput.

In summary, our key contributions are:
\begin{itemize}[noitemsep,nolistsep]
    \item A formulation of heterogeneity-aware multi-LLM serving as a joint optimization over resource allocation and model placement under per-model latency SLOs.
    \item A lossless decomposition that shifts the slow model placement search offline via reusable \emph{Serving Templates}, reducing online solve times to tens of seconds.
    \item ILP-based solvers for both offline Serving Template generation and online resource allocation that scale to clusters of hundreds of nodes.
    \item An end-to-end distributed implementation of \sys, alongside a high-fidelity simulator validated against real hardware for large-scale evaluation.
    \item Comprehensive evaluations across diverse models, workloads, and cluster scales, demonstrating up to 2.79$\times$ in cost savings and 2.39$\times$ higher goodput.
\end{itemize}
We will open-source the system upon paper acceptance.

\section{Background and Motivation}
\label{sec:background}

\subsection{Distributed LLM Serving}

\myparagraph{Hardware Affinity in LLM Inference.}
Modern LLMs use diverse Transformer architectures, including full-attention~\cite{grattafiori2024llama}, hybrid-attention with sparse patterns for long contexts (e.g., sliding-window attention)~\cite{agarwal2025gpt,gemma3}, and mixture-of-experts (MoE)~\cite{fedus2022switch,jiang2024mixtral,agarwal2025gpt,yang2025qwen3}. These designs fundamentally alter the balance between computation, memory capacity, and bandwidth requirements. For instance, MoE activates a sparse expert subset per token, decoupling model size from FLOPs~\cite{fedus2022switch,jiang2024mixtral}. Furthermore, LLM serving operates in two distinct phases: prefill (prompt processing) and decode (auto-regressive generation)~\cite{yu2022orca}. These phases exhibit contrasting hardware demands: prefill is highly parallel and compute-bound, whereas decode is sequential and heavily bottlenecked by memory bandwidth, particularly for KV-cache accesses~\cite{zhong2024distserve,patel2024splitwise}. Consequently, optimal GPU selection depends jointly on model architecture and serving phase rather than a uniform hardware assignment~\cite{mei2025helix,zhong2024distserve,patel2024splitwise}.

\myparagraph{Parallelism Strategies for Heterogeneous LLM Serving.}
Distributed LLM serving relies on four primary parallelism strategies: data (DP)~\cite{dean2012large}, pipeline (PP)~\cite{huang2019gpipe}, tensor (TP)~\cite{shoeybi2019megatron}, and expert (EP)~\cite{lepikhin2020gshard}. DP replicates models to distribute request load, while PP partitions the model into sequential layer stages~\cite{dean2012large,huang2019gpipe}. In contrast, TP shards individual operators across devices, and EP distributes MoE experts, routing tokens dynamically~\cite{shoeybi2019megatron,lepikhin2020gshard}. DP and PP serve as effective building blocks for heterogeneous serving due to their coarse-grained partitioning: DP distributes requests proportionally, and PP adapts to varying device capabilities via uneven stage sizing~\cite{mei2025helix}. However, TP and EP require fine-grained, tightly synchronized communication—such as all-reduce in TP and all-to-all exchanges in EP~\cite{shoeybi2019megatron,lepikhin2020gshard}. This tight coupling makes them highly sensitive to compute and network imbalances, rendering them poorly suited for heterogeneous clusters.


\subsection{Challenges and Opportunities}
\label{sec:opportunities}

\myparagraph{GPU Heterogeneity.} Serving a single model replica across heterogeneous GPUs tightly couples model placement and resource allocation, creating an exponential search space. However, this complexity unlocks critical cost and performance opportunities.

Fig.~\ref{fig:sec2_challenges_mixed_slo} illustrates the most cost-efficient prefill strategy for Qwen-3 235B~\cite{yang2025qwen3} under an 1800~ms latency SLO, using the five GPU types in Table~\ref{tab:gpu-specs}. The optimal configuration is a mixed pipeline of three single-GPU L40S nodes and three dual-GPU H100 nodes with a non-uniform layer partition. Notably, this outperforms all homogeneous H100 setups (526 vs.\ 481, 460, and 476 Tok/s/USD). Thus, using mid-tier GPUs not only mitigates top-tier scarcity but actively \emph{improves} cost efficiency—a pattern consistent across other models and GPU combinations. Furthermore, heterogeneous pipelines enable finer-grained capacity scaling. As Fig.~\ref{fig:sec2_challenges_mixed_throughput} shows, homogeneous node sets (gray) produce discrete throughput tiers, forcing allocators to over-provision when demand falls between steps. In contrast, heterogeneous node sets (orange) generate a near-continuous throughput spectrum. This allows the system to tightly match per-model demand and eliminate wasted capacity. Together, these two effects drive the up to 2.79$\times$ cost reduction reported in Sec.~\ref{sec:evaluation}.

\myparagraph{Joint Optimization across LLMs.} The second challenge is coordinating decisions \emph{across} models rather than solving them in isolation. Fig.~\ref{fig:sec2_challenges_multi_llm} demonstrates the limitations of a per-model greedy strategy. When two models (M1, M2) share a constrained pool (2 GPU-A, 3 GPU-B nodes), independent allocation (\circledone) causes M1 to monopolize the GPU-A nodes and over-serve its demand, leaving M2 under-resourced with only GPU-B nodes. Joint optimization (\circledtwo) resolves this contention: M1 accepts a mixed pipeline (one GPU-A, one GPU-B node), freeing a GPU-A node for M2 to form its own mixed replica. Consequently, both models meet their throughput demands without idle resources. Capturing these global efficiency gains requires jointly optimizing allocation and placement across \emph{all} models—the core challenge \sys addresses.


\begin{figure*}[t]
    \centering
    \includegraphics[width=0.9\linewidth]{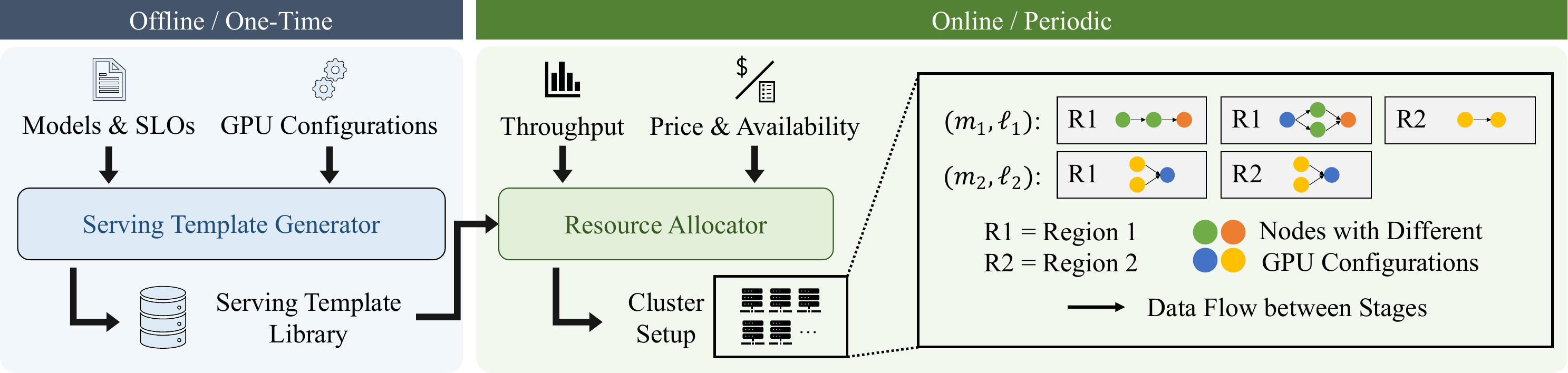}
    \caption{The two-stage workflow of \sys. \emph{Offline}, the Serving Template generator (Sec.~\ref{sec:template_generation}) pre-computes a template library from target models, SLOs, and GPU configurations. \emph{Online}, the resource allocator (Sec.~\ref{sec:resource_allocation}) queries this library to dynamically provision a cost-minimal, cross-region cluster that satisfies throughput demands. The offline stage runs once, enabling the online allocator to periodically adapt to workload and resource fluctuations.}
    \label{fig:sec3_workflow}
    \Description{Two-stage workflow of \sys.}
\end{figure*}

\section{The Multi-LLM Serving Problem}
\label{sec:problem_formulation}


We formulate the multi-LLM serving problem on heterogeneous GPUs as a joint optimization over resource allocation and model placement. Let $\mathcal{G}$ denote the pool of available GPU nodes, where each node $g \in \mathcal{G}$ incurs a provisioning cost $p(g) \ge 0$. Let $\mathcal{M}$ be the set of models, with each model $m \in \mathcal{M}$ requiring a throughput $T_m$ under its latency SLO.

A serving strategy is defined by two functions. First, the resource allocation $\Phi$ determines a replica count $n_m \ge 1$ for each model $m$ and assigns each replica a disjoint subset of nodes:
\[
  \begin{gathered}
    \Phi(m) = \bigl(G_m^{1}, \dots, G_m^{n_m}\bigr),
    \quad G_m^{i} \subseteq \mathcal{G}, \\
    G_m^{i} \cap G_{m'}^{j} = \varnothing \text{ for } (m,i) \neq (m',j).
  \end{gathered}
\]
Second, for each replica $G_m^{i}$, the model placement $\Psi$ defines a hybrid pipeline- and data-parallel layout. It specifies the number of pipeline stages, the partition of layers across stages, and the node-to-stage mapping. Multiple nodes mapped to the same stage duplicate its layers to share the load (DP). Following prior work~\cite{mei2025helix, jiang2023hexgen}, we restrict inter-node parallelism to PP and DP; tensor and expert parallelism shard work symmetrically and bottleneck at the slowest device, rendering them inefficient across heterogeneous nodes. However, TP and EP remain active intra-node, where GPUs are homogeneous and share high-bandwidth interconnects.

Let $T(\Psi(G_m^{i}))$ denote the throughput of model $m$'s $i$-th replica under $\Psi$, subject to its latency SLO. The system jointly optimizes $\Phi$ and $\Psi$ to minimize total provisioning cost while satisfying per-model throughput and latency demands:
\begin{alignat*}{2}
  \min_{\Phi,\, \Psi} \quad
    & \sum_{m \in \mathcal{M}} \sum_{i=1}^{n_m}
      \sum_{g \in G_m^{i}} p(g) && \\[-1pt]
  \text{s.t.} \quad
    & \sum_{i=1}^{n_m} T\!\bigl(\Psi(G_m^{i})\bigr) \;\ge\; T_m,
      \quad && \forall\, m \in \mathcal{M}.
\end{alignat*}


\section{Optimization Formulation in \sys}
\label{sec:solution}


\subsection{Solution Overview}
\label{sec:solution_overview}

\myparagraph{Naive Enumeration is Intractable.}
Optimizing resource allocation $(\Phi)$ and model placement $(\Psi)$ jointly via naive enumeration is intractable, as both subproblems exhibit exponential search spaces. At scale, this combinatorial explosion is severe. For example, Helix~\cite{mei2025helix} requires a four-hour search to optimize placement for a single model on just 24 nodes, even \emph{without} a latency SLO. Enforcing an SLO further compounds this complexity, as latency feasibility can only be verified after a full placement is finalized. Consequently, computing even a single global solution is computationally prohibitive, precluding the system from dynamically adapting to fluctuating demands and resource availability online.

\myparagraph{Key Insight.}
To enable rapid online reconfiguration, \sys exploits a fundamental substructure of the joint problem: for a fixed model and latency SLO, the throughput-optimal placement $\Psi^*(\mathcal{G}')$ on a node subset $\mathcal{G}'$ depends solely on $\mathcal{G}'$, independent of the broader cluster allocation. This decoupling allows \sys to pre-compute and cache throughput-optimal placements for distinct node combinations offline. We term these reusable artifacts \textit{Serving Templates}. While caching all combinations yields a theoretically lossless decomposition, the actual space is infinite. Therefore, \sys limits enumeration to a principled subset of combinations (Sec.~\ref{sec:template_generation}), which we demonstrate has a negligible performance impact (Sec.~\ref{sec:sensitivity}). Crucially, this decomposition eliminates the placement search from the online critical path.

\myparagraph{Two-Stage Workflow.} 
Building on this insight, \sys orchestrates multi-LLM serving in two distinct stages (Fig.~\ref{fig:sec3_workflow}). \emph{Offline}, the Serving Template generator (Sec.~\ref{sec:template_generation}) enumerates node combinations for each model and SLO pair $(m, \ell)$. It employs integer linear programming (ILP) to discover the throughput-optimal placement for each node combination, populating a \emph{Serving Template Library}. \emph{Online}, the resource allocator (Sec.~\ref{sec:resource_allocation}) queries this library with a lightweight ILP to rapidly select and instantiate templates across regions, fulfilling global throughput demands at minimal cost while respecting current resource constraints. By moving the heavy placement search offline, online allocation resolves in tens of seconds, ensuring agile cluster adaptation.


\subsection{Serving Template Generation}
\label{sec:template_generation}

\begin{figure}
    \centering
    \includegraphics[width=0.9\linewidth]{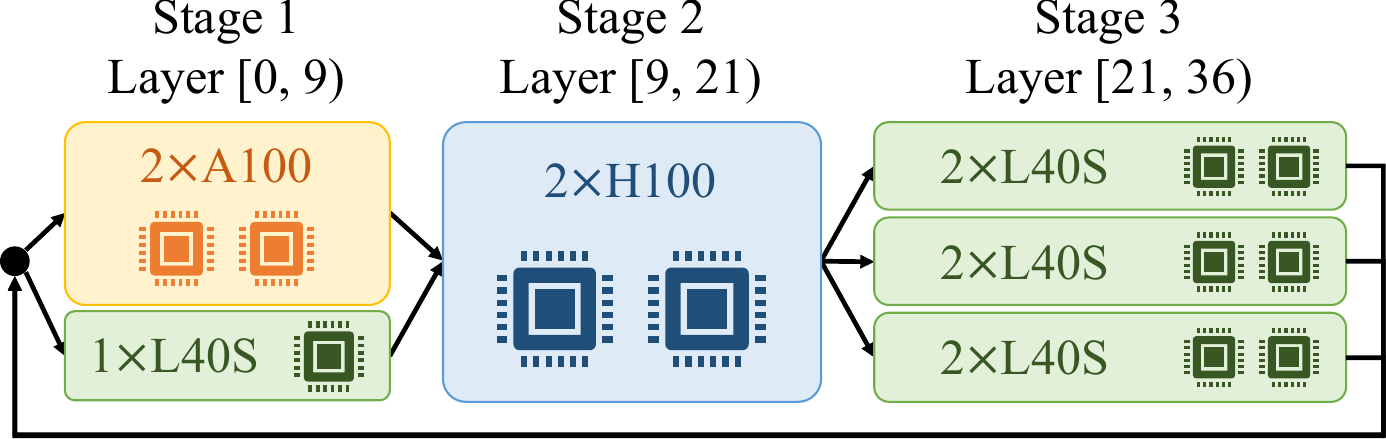}
    \caption{\textbf{Serving Template.} Boxes denote nodes; arrows indicate data flow. Nodes within the same pipeline stage replicate identical model layers, functioning as data-parallel replicas. During execution, a request traverses a single node per stage, looping back for auto-regressive generation.}
    \label{fig:sec3_serving_template}
    \Description{Illustration of a Serving Template.}
\end{figure}

\myparagraph{Serving Template.}
Formally, a Serving Template for model $m$ under latency SLO $\ell$ is a tuple $\tau=(m, \ell, \mathcal{G}', \Psi^*(\mathcal{G}'))$. Here, $\mathcal{G}'$ is a set of nodes within a single region. Templates are strictly intra-region because inter-region network latencies (10 - 100+ milliseconds) prohibit efficient pipeline parallelism. Templates are uniquely identified by their multiset of GPU configurations. As \sys employs prefill-decode disaggregation, we generate separate templates for each phase, mapping $\ell$ to time-to-first-token (TTFT) or time-per-output-token (TPOT), and assuming $\ell$ is a fixed, known workload characteristic. $\Psi^*(\mathcal{G}')$ represents the throughput-optimal model placement for $m$ on $\mathcal{G}'$ subject to $\ell$, yielding the template's throughput $T(\tau) := T(\Psi^*(\mathcal{G}'))$. Fig.~\ref{fig:sec3_serving_template} illustrates a template combining 6 nodes across 4 GPU configurations into a 3-stage pipeline with intra-stage data parallelism.

\myparagraph{Optimal Model Placement via ILP.}
For given node set $\mathcal{G}'$, model $m$ with $L$ layers, and SLO $\ell$, we formulate the search for $\Psi^{*}(\mathcal{G}')$ as an ILP parameterized by pipeline depth $S$.

\textit{Decision variables.} We define two sets of binary variables: $x_{sj}=1$ iff stage $s \in [1, S]$ holds $j \in [1, L]$ consecutive layers, and $y_{sk}=1$ iff node $g_k \in \mathcal{G}'$ is assigned to stage $s$.

\textit{Constraints.} Each stage selects exactly one layer count, and each node maps to exactly one stage. The stage layer counts must sum to the full model size $L$:
$$\sum_{j=1}^{L} x_{sj} = 1 \ \forall s, \quad \sum_{s=1}^{S} y_{sk} = 1 \ \forall k, \quad \sum_{s=1}^{S} \sum_{j=1}^{L} j \cdot x_{sj} = L.$$

Throughput is bottlenecked by the slowest pipeline stage. Because nodes assigned to the same stage run data parallelism, their throughputs aggregate. For each stage $s$:
\[
T(\tau) \;\le\; \sum_{j=1}^{L} \sum_{k=1}^{|\mathcal{G}'|} x_{sj} \cdot y_{sk} \cdot \hat{T}_{j}(g_k)
\]
where $\hat{T}_{j}(g_k)$ is the maximum throughput of node $g_k$ hosting $j$ layers under a per-stage latency budget of $\ell / S$. This budget derives from Little's Law: for a saturated pipeline with $S$ stages and continuous micro-batching, end-to-end latency is bounded by $S$ times the slowest stage's latency. We determine $\hat{T}_{j}(g_k)$ via a one-time offline profiling step for each GPU configuration, maximizing batch size within memory limits using statistical mean input and output sequence lengths.

\textit{Linearization.} To eliminate the quadratic term $x_{sj} \cdot y_{sk}$, we introduce a binary auxiliary variable $z_{sjk}$ subject to standard McCormick envelope constraints:
$$z_{sjk} \le x_{sj}, \qquad z_{sjk} \le y_{sk}, \qquad z_{sjk} \ge x_{sj} + y_{sk} - 1.$$
The throughput constraint then becomes linear:
\[
T(\tau) \;\le\; \sum_{j=1}^{L} \sum_{k=1}^{|\mathcal{G}'|} z_{sjk} \cdot \hat{T}_{j}(g_k)
\]

\textit{Objective.} The ILP maximizes the end-to-end throughput $T(\tau)$. Solving this ILP for a given $S$ yields the optimal layer partition ($x_{sj}$) and node-to-stage assignment ($y_{sk}$). The global throughput-optimal placement $\Psi^{*}(\mathcal{G}')$ is the configuration achieving the highest throughput across all valid pipeline depths $S \in [1, |\mathcal{G}'|]$.

This formulation contains $\mathcal{O}(S \cdot L \cdot |\mathcal{G}'|)$ binary variables and constraints. Using modern solvers like Gurobi~\cite{gurobi}, evaluating a single $S$ takes seconds, making a full enumeration over $S$ highly efficient.

\myparagraph{Generating the Serving Template Library.}
While preserving the original search space theoretically requires evaluating all possible node combinations, the unbounded nature of this space necessitates pruning. \sys restricts enumeration to combinations containing at most $N_{\max}$ nodes and a total GPU memory capacity below $\rho$ times the model size. Combinations exceeding these thresholds incur high communication overheads and are typically dominated by splitting the same resources into smaller, independent replicas. Consequently, the resource allocator rarely selects them. As shown in Sec.~\ref{sec:sensitivity}, this pruning maintains near-optimal cost efficiency. 
The resulting library contains thousands to tens of thousands of templates per $(m, \ell)$ pair. Because each ILP solves independently, library generation is embarrassingly parallel, completing in minutes for small models and tens of minutes for larger ones (Sec.~\ref{sec:sensitivity}).


\subsection{Online Resource Allocation}
\label{sec:resource_allocation}

\myparagraph{Serving Instance.}
To bridge offline templates and online execution, we define a \emph{Serving Instance} as the physical realization of a template within a specific region. Each instance provisions the exact node combination $\mathcal{G}'$ dictated by its template and deploys a single model replica using the pre-computed optimal layout $\Psi^*(\mathcal{G}')$. By treating Serving Instances as atomic building blocks, \sys decouples model placement from resource allocation, reducing online problem to selecting which templates to instantiate and where.

\begin{figure}
    \centering
    \includegraphics[width=\linewidth]{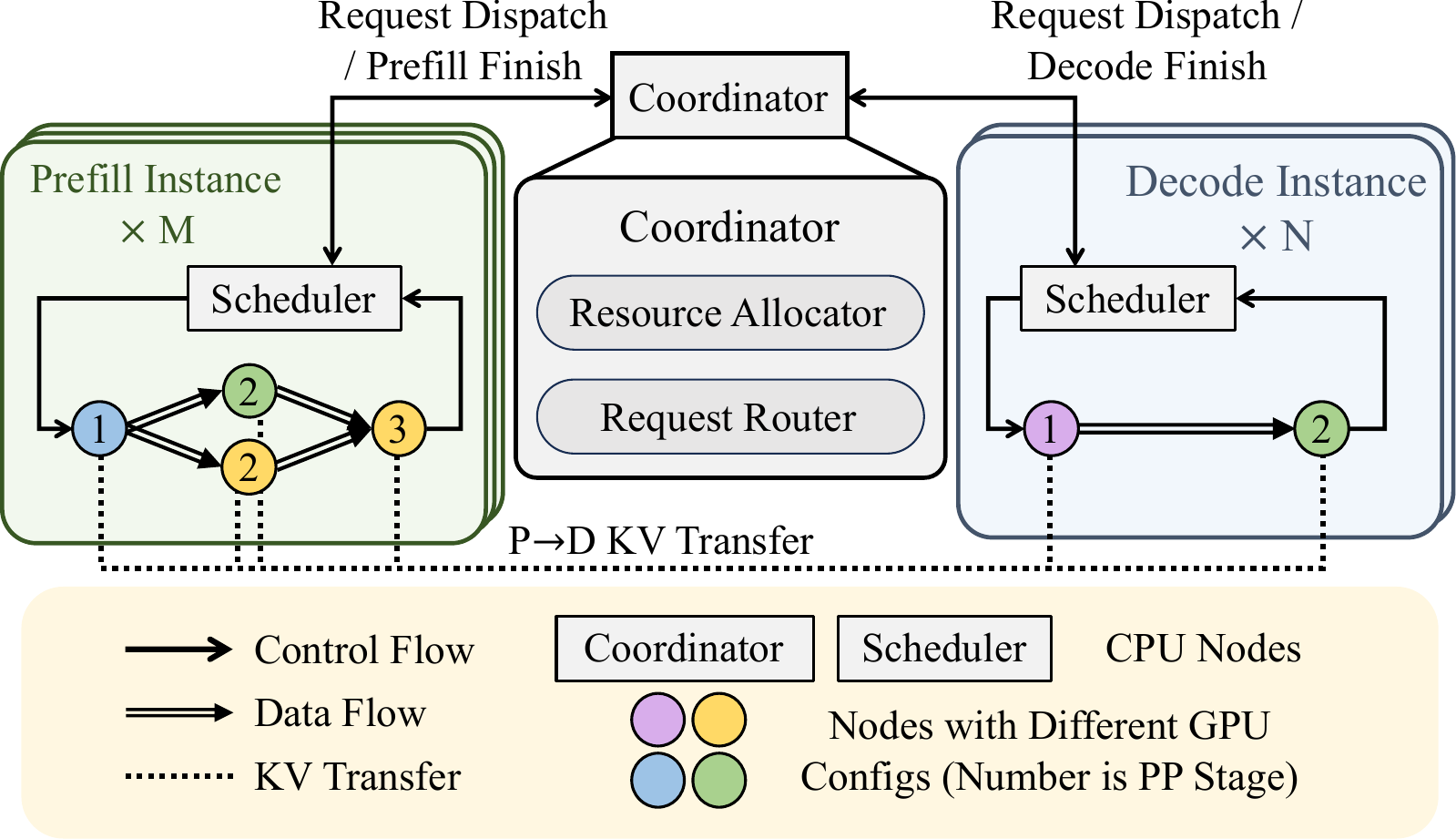}
    \caption{\textbf{\sys runtime architecture.} A central coordinator manages resource allocation and routes requests across prefill and decode Serving Instances. Intra-instance schedulers dispatch these requests across heterogeneous engine nodes (colored circles) per pipeline stage. KV caches transfer peer-to-peer from prefill to decode nodes.}
    \label{fig:sec4_runtime}
    \Description{Overview of the \sys runtime system.}
\end{figure}

\myparagraph{Optimal Resource Allocation via ILP.}
Let $\mathcal{R}$ denote the available cloud regions and $\mathcal{C}$ the GPU configurations. For each region $r \in \mathcal{R}$ and configuration $c \in \mathcal{C}$, let $A_r(c) \ge 0$ be the available node count, and $p_r(c)$ the per-node provisioning cost. For a model $m \in \mathcal{M}$ requiring throughput $T_m$ under SLO $\ell_m$, the offline stage supplies $N_m$ valid templates. Let $\tau^m_i$ denote the $i$-th template, $T(\tau^m_i)$ its throughput, and $U_c(\tau^m_i)$ its required number of type-$c$ nodes. The total provisioning cost of $\tau^m_i$ in region $r$ is $p_r(\tau^m_i) := \sum_{c \in \mathcal{C}} p_r(c) \cdot U_c(\tau^m_i)$.

\textit{Decision variables.} An integer variable $\nu_r(\tau^m_i) \ge 0$ dictates the number of type $\tau^m_i$ instances to deploy in region $r$.

\textit{Constraints.} Aggregate node consumption cannot exceed regional availability for any GPU configuration $c$:
\[
  \sum_{m \in \mathcal{M}} \sum_{i=1}^{N_m} U_c(\tau^m_i) \cdot \nu_r(\tau^m_i)
  \;\le\; A_r(c) \quad \forall r, c.
\]
For each model $m$, total throughput must satisfy its demand:
\[
  \sum_{r \in \mathcal{R}} \sum_{i=1}^{N_m} T(\tau^m_i) \cdot \nu_r(\tau^m_i)
  \;\ge\; T_m \quad \forall m.
\]

\textit{Objective.} While Sec.~\ref{sec:problem_formulation} targets a static snapshot, the online allocator dynamically steers the cluster. Adding instances incurs non-trivial overhead (e.g., node startup, weight loading), whereas tear-down is graceful since the runtime transparently drains in-flight requests (Sec.~\ref{sec:system_design}). To penalize churn, we apply an initialization cost only to \emph{newly added} instances. Let $\nu'_r(\tau^m_i)$ be the current number of running instances (a constant at solve time), and $I_r(\tau^m_i) \ge 0$ a continuous auxiliary variable bounded by:
\[
  I_r(\tau^m_i) \;\ge\;
    \bigl(\nu_r(\tau^m_i) - \nu'_r(\tau^m_i)\bigr) \cdot p_r(\tau^m_i) \cdot K,
\]
where $K$ is the ratio of node initialization time to the cluster adjustment interval. Because $I_r(\tau^m_i) \ge 0$, the bound is active only when scaling up ($\nu_r > \nu'_r$), meaning scaling down contributes nothing to the objective. The allocator minimizes the total provisioning cost plus initialization penalties:
\[
  \min \;\; \sum_{r \in \mathcal{R}} \sum_{m \in \mathcal{M}} \sum_{i=1}^{N_m}
    \Bigl[\, \nu_r(\tau^m_i) \cdot p_r(\tau^m_i) + I_r(\tau^m_i) \,\Bigr].
\]

\myparagraph{Tractability.}
This ILP introduces $2|\mathcal{R}| \cdot \sum_{m\in \mathcal{M}} N_m$ integer and continuous variables, subject to $|\mathcal{C}| \cdot |\mathcal{R}| + |\mathcal{M}|$ physical constraints and $|\mathcal{R}| \cdot \sum_{m\in \mathcal{M}} N_m$ penalty bounds. Although variable counts can reach the millions for large template libraries, the optimal solution is exceptionally sparse: a cluster of hundreds of nodes typically activates only tens of templates. Consequently, the vast majority of $\nu_r(\tau^m_i)$ variables evaluate to zero, trivially satisfying their penalty bounds. As demonstrated in Sec.~\ref{sec:evaluation}, commercial solvers like Gurobi~\cite{gurobi} exploit this sparsity to resolve the ILP in tens of seconds, safely within our online scheduling budget.

\section{\sys Runtime}
\label{sec:runtime}

\subsection{System Design}
\label{sec:system_design}

Fig.~\ref{fig:sec4_runtime} illustrates \sys's runtime architecture, comprising a central CPU \emph{coordinator node} alongside pools of prefill and decode \emph{Serving Instances}. The coordinator hosts the resource allocator (Sec.~\ref{sec:resource_allocation}) and a request router. Each Serving Instance encapsulates a CPU \emph{scheduler node}—managing intra-instance scheduling and coordinator communication—and a group of GPU \emph{engine nodes} that execute model inference according to the template's model placement. While we describe \sys under prefill-decode (PD) disaggregation, this architecture natively supports PD-aggregated serving.

\myparagraph{Request Life-Cycle.}
Upon request arrival, the coordinator routes it to a prefill instance via weighted round-robin, proportional to the instance's template throughput $T(\tau)$. The instance's scheduler then assigns exactly one engine node per pipeline stage, again using weighted round-robin based on each node's expected throughput under the model placement. The scheduler dispatches the prompt to the first stage, and engine nodes stream intermediate activations down the pipeline. Upon prefill completion, the scheduler notifies the coordinator, which selects a decode instance. The respective schedulers then orchestrate a direct, peer-to-peer transfer of the KV cache. Once transferred, the decode instance starts auto-regressive generation.

\myparagraph{Instance Life-Cycle.}
\sys invokes the allocator periodically to reconcile the running cluster against dynamic throughput demands, resource availability, and hardware pricing. 

\textit{Scaling down.} When a template's target count decreases, \sys gracefully terminates the excess instances, prioritizing those with the lowest active load. A terminating instance ceases accepting new requests and shuts down only after its in-flight requests complete (\emph{connection draining}). Because this draining typically finishes within minutes, it is preferable to request migration; migrating in-flight decode requests requires transferring massive KV caches across the network, severely degrading system efficiency.

\textit{Scaling up.} To scale up a template, \sys provisions new nodes and initializes instances according to the template's exact model placement. Dominated by node startup, weight loading, and CUDA graph compilation, this process takes several minutes—an overhead explicitly modeled by the allocator's initialization penalty (Sec.~\ref{sec:resource_allocation}). Once fully initialized, the scheduler registers with the coordinator, and the router begins dispatching requests to the new instance.


\subsection{Implementation}
\label{sec:implementation}

We implement the \sys runtime in 53K LoC of Python and C++. The control plane runs over ZeroMQ~\cite{hintjens2013zeromq}, on top of which we build Serving Instance life-cycle management and request dispatching. For intra-instance communication between engine nodes, we build a high-bandwidth, low-latency pipeline-parallel framework on top of NCCL~\cite{nvidia_nccl}. This framework supports arbitrary pipeline- and data-parallel configurations and transparently enables Remote Direct Memory Access (RDMA) and GPUDirect RDMA~\cite{nvidia_gpudirect} whenever the hardware permits. For KV cache transfers between prefill and decode instances, we use Gloo~\cite{meta_gloo} over CPU-based RDMA. Routing the transfer through the CPU isolates KV cache movement from the GPUs, ensuring it does not contend with model inference or pipeline-parallel communication for GPU resources. We use vLLM~\cite{kwon2023efficient} as the per-node execution engine; however, the runtime is engine-agnostic, and any compatible inference engine can be substituted without changes to the surrounding system.

To evaluate \sys at scales of hundreds of nodes---where physical deployment is cost-prohibitive---we developed an event-based simulator in 5K LoC of Python. Driven by empirical offline profiling data from our heterogeneous GPU pool, the simulator models execution at the granularity of individual pipeline stages per engine node. As Sec.~\ref{sec:simulator_fidelity} demonstrates, it reproduces real-world per-request prefill and decode latencies to within 5.6\% and 7.2\% respectively, confirming its high fidelity as a proxy for large-scale experiments.

\section{Evaluation}
\label{sec:evaluation}

We evaluate \sys to answer the following questions:
\begin{itemize}[noitemsep,nolistsep]
    \item Does \sys reduce serving costs across diverse model mixtures and heterogeneous GPU pools? (Sec.~\ref{sec:e2e_exp1})
    \item Can \sys sustain high goodput and mitigate contention under tight resource availability? (Sec.~\ref{sec:e2e_exp2})
    \item How resilient are \sys's cost savings to varying throughput distributions across models? (Sec.~\ref{sec:e2e_exp3})
    \item Does \sys outperform SOTA heterogeneous serving (Helix) even in the single-model regime? (Sec.~\ref{sec:helix_comparison})
\end{itemize}


\subsection{Experiment Setup}
\label{sec:experiment_setup}

\myparagraph{Model and GPU Setup.} We evaluate \sys across two deployment scales. The \smallscale features three models (Qwen-3 32B~\cite{yang2025qwen3}, GPT-OSS 20B~\cite{agarwal2025gpt}, Phi4-14B~\cite{abdin2024phi}) running on 12 GPU configurations (1/2/4/8$\times$ L40S, L4, A10G) across two cloud regions, requiring 20--40 GPUs. The \largescale \textbf{adds} three larger models (Qwen-3 235B~\cite{yang2025qwen3}, GPT-OSS 120B~\cite{agarwal2025gpt}, Llama-3 70B~\cite{grattafiori2024llama}) and eight top-tier configurations (1/2/4/8$\times$ H100, A100), expanding to a third region and 100--300 GPUs total. Table~\ref{tab:model_info} details model properties. We evaluate the \smallscale on real hardware (Sec.~\ref{sec:e2e_exp1}, \ref{sec:e2e_exp2}) and in simulation (Sec.~\ref{sec:e2e_exp3}). The \largescale runs exclusively in simulation to avoid the prohibitive (thousands of dollars per hour) cost of provisioning hundreds of GPUs. Re-running the real system experiments in simulation (Sec.~\ref{sec:simulator_fidelity}) yields prefill and decode latency errors of only 5.6\% and 7.2\% on average, confirming the simulator's high fidelity.

\myparagraph{Workload Setup.} Request lengths and arrival patterns are drawn from Azure Code~\cite{stojkovic2025dynamollm}, Azure Conversation~\cite{stojkovic2025dynamollm}, and BurstGPT~\cite{wang2025burstgpt}. We assign datasets evenly across models. Table~\ref{tab:model_info} maps datasets and latency SLOs per model; SLOs are calibrated to model size and architecture (e.g., MoE~\cite{fedus2022switch} vs.\ hybrid attention~\cite{beltagy2020longformer}) following AdaServe~\cite{li2025adaserve}. We scale traces uniformly to achieve a default average arrival rate of 10~req/s (\smallscale) or 25~req/s (\largescale) per model. Sec.~\ref{sec:e2e_exp3} evaluates skewed arrival rates, routing 80\% of traffic strictly to either large or small models.

\begin{table}[t]
\centering
\footnotesize
\setlength{\tabcolsep}{4pt}
\renewcommand{\arraystretch}{1.15}
\begin{tabular}{@{}l|c|c|c|c|c|c@{}}
\toprule
\makecell{\textbf{Model}}
  & \makecell{\textbf{MoE}}
  & \makecell{\textbf{Hyb.}\\\textbf{Attn.}}
  & \makecell{\textbf{$L$}}
  & \makecell{\textbf{Prefill}\\\textbf{(ms)}}
  & \makecell{\textbf{Decode}\\\textbf{(ms)}}
  & \makecell{\textbf{Trace}} \\
\midrule
Phi4 14B     & \no  & \no  & 40 & 1200 & 60  & AzureConv. \\
GPT-OSS 20B  & \yes & \yes & 24 & 900  & 30  & AzureCode  \\
Qwen3 32B    & \no  & \no  & 64 & 1600 & 100 & BurstGPT   \\
Llama3 70B   & \no  & \no  & 80 & 1500 & 80  & BurstGPT   \\
GPT-OSS 120B & \yes & \yes & 36 & 1000 & 40  & AzureConv. \\
Qwen3 235B   & \yes & \no  & 94 & 1800 & 120 & AzureCode  \\
\bottomrule
\end{tabular}
\caption{\textbf{Model characteristics and serving metrics.} First three columns: the use of mixture-of-experts (MoE)~\cite{fedus2022switch}, hybrid sliding-window attention, and total layer count ($L$).}
\vspace{-2em}
\label{tab:model_info}
\end{table}
\begin{figure}
    \centering
    \includegraphics[width=\linewidth]{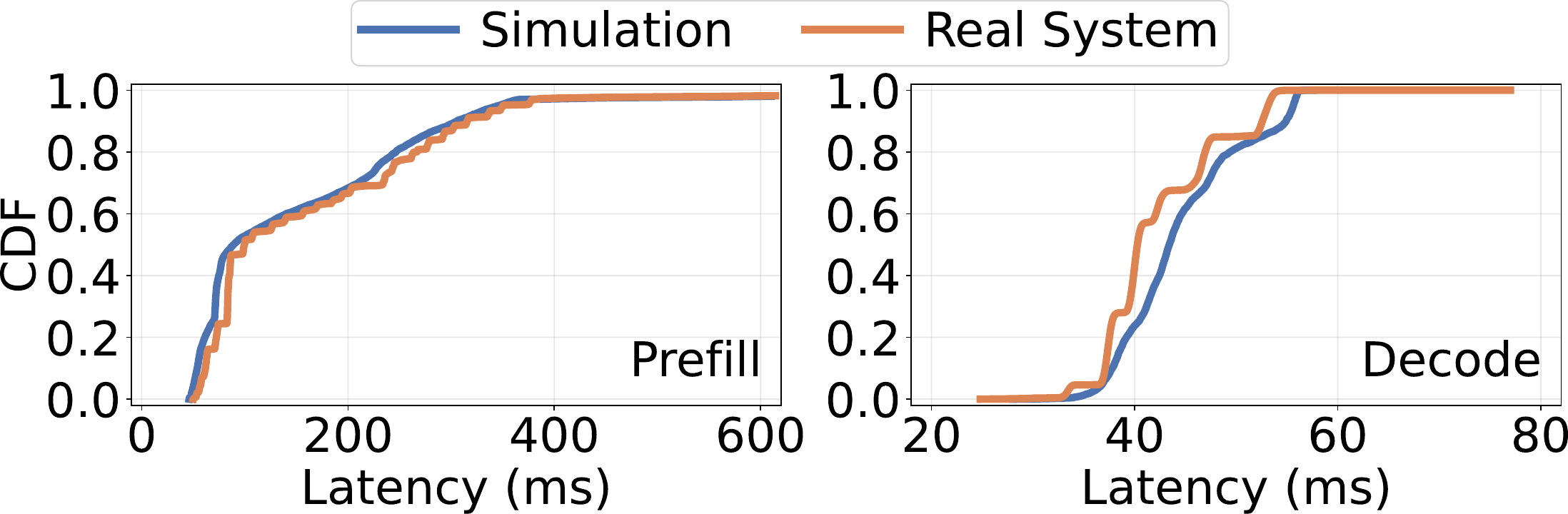}
    \caption{Simulated prefill and decode latency CDFs for Phi-4 14B tightly align with real-hardware measurements.}
    \label{fig:simulator_fidelity}
    \Description{Simulator fidelity.}
\end{figure}
\begin{figure*}
    \centering
    \begin{subfigure}[t]{0.2262\linewidth}
        \centering
        \includegraphics[width=\linewidth]{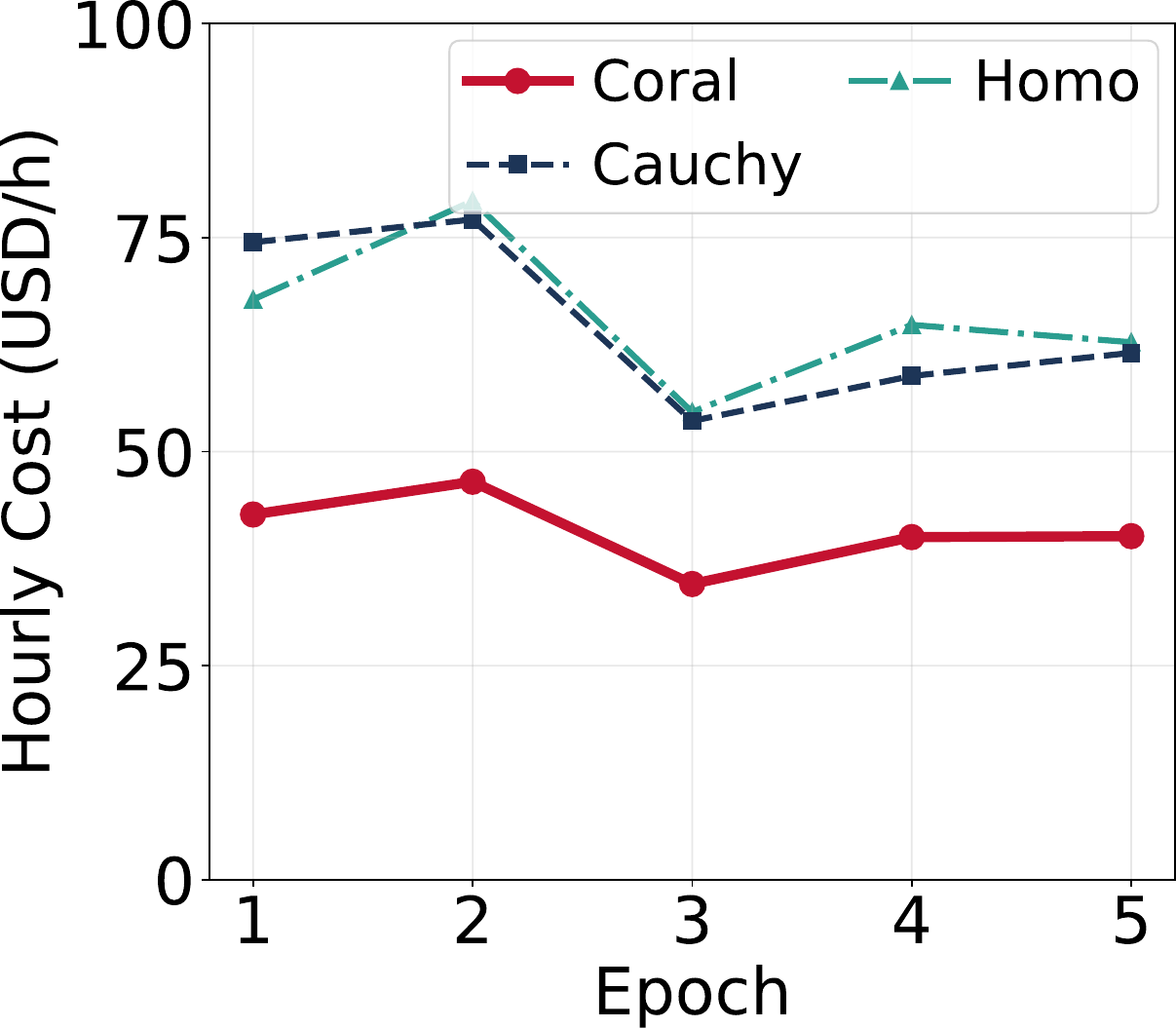}
        \caption{Hourly cost comparison in \smallscale.}
        \label{fig:e2e_exp1_a}
    \end{subfigure}
    \hfill
    \begin{subfigure}[t]{0.2262\linewidth}
        \centering
        \includegraphics[width=\linewidth]{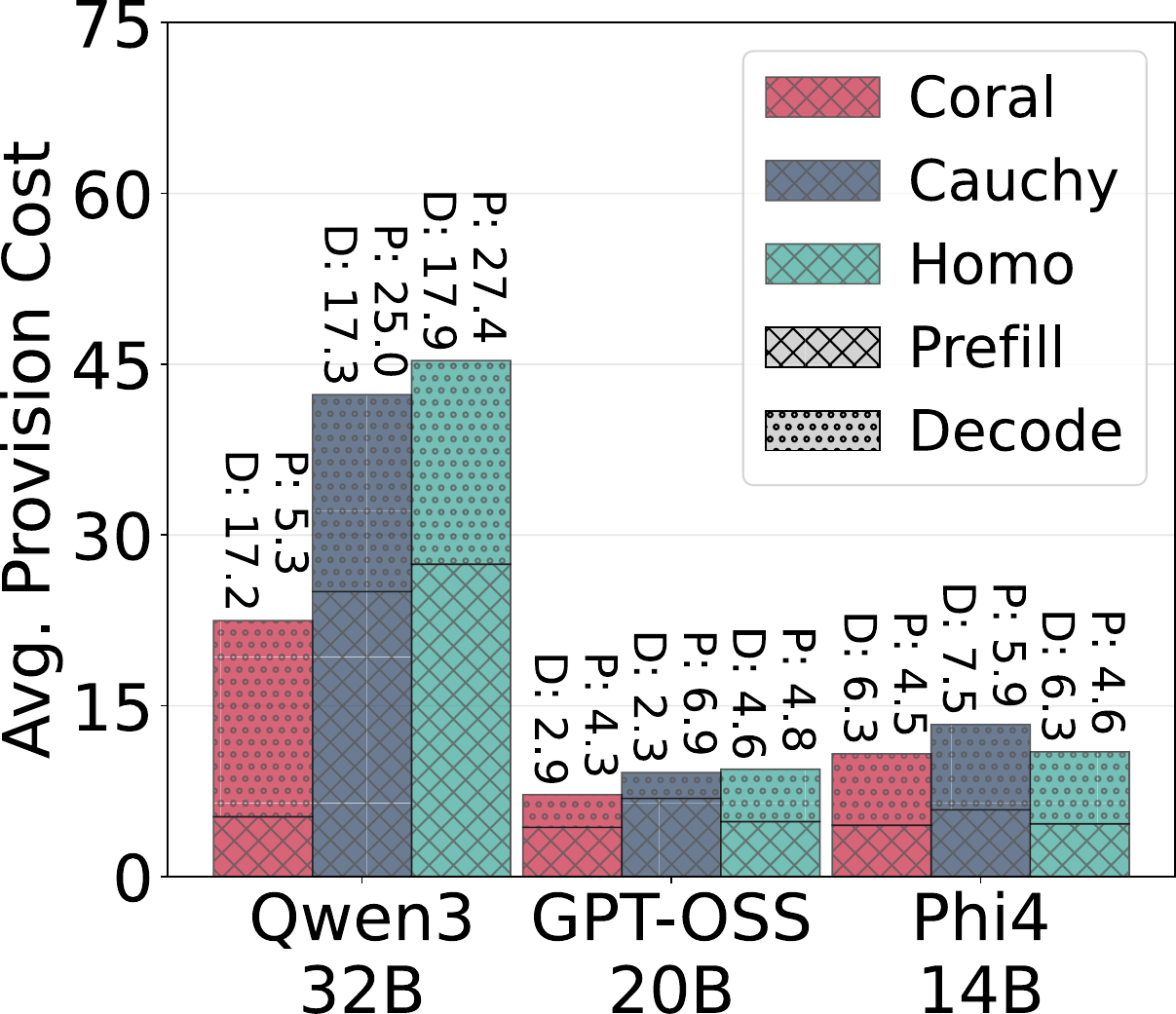}
        \caption{Per-model average provision cost breakdown in \smallscale.}
        \label{fig:e2e_exp1_b}
    \end{subfigure}
    \hfill
    \begin{subfigure}[t]{0.2262\linewidth}
        \centering
        \includegraphics[width=\linewidth]{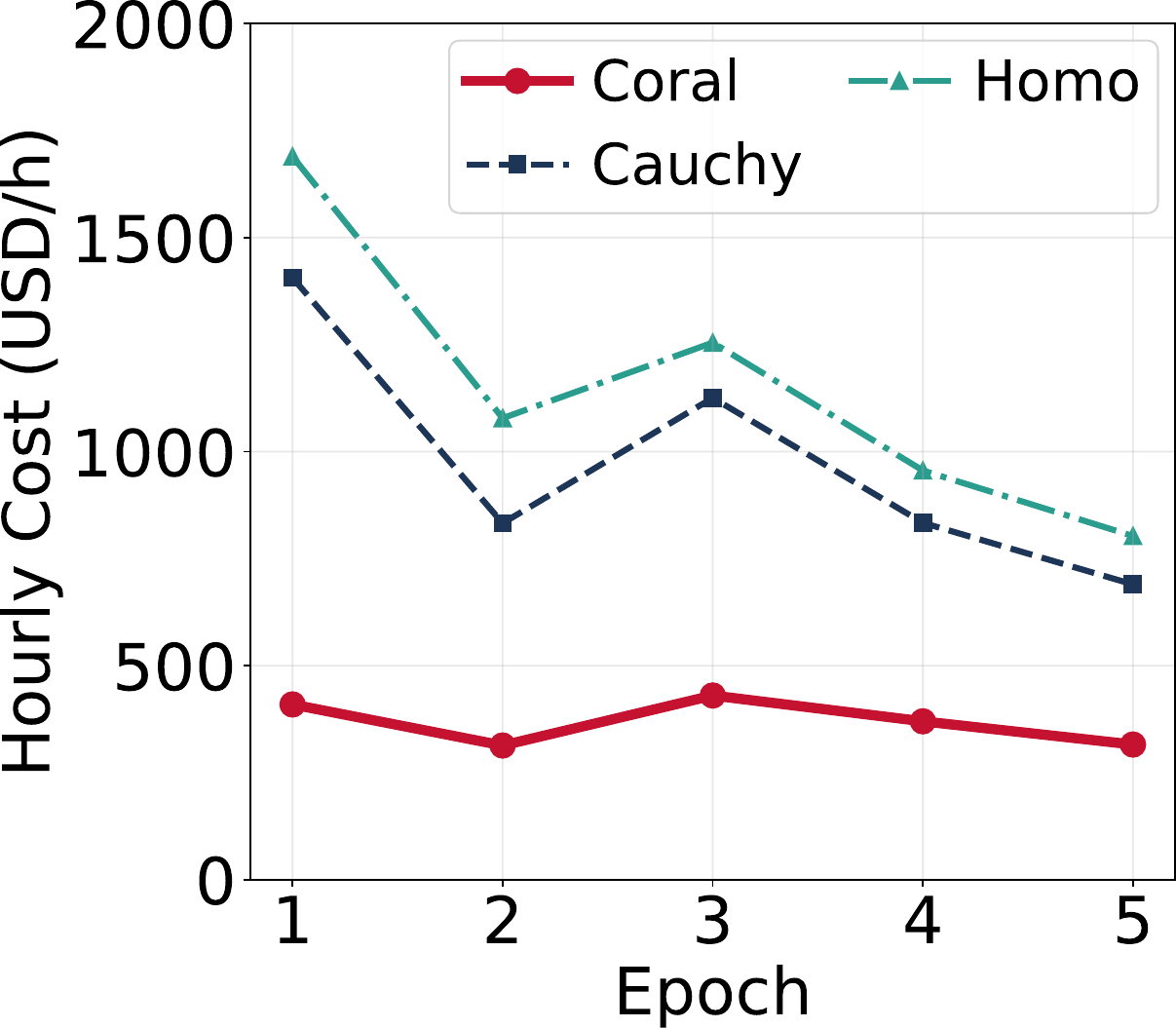}
        \caption{Hourly cost comparison in \largescale.}
        \label{fig:e2e_exp1_c}
    \end{subfigure}
    \hfill
    \begin{subfigure}[t]{0.2817\linewidth}
        \centering
        \includegraphics[width=\linewidth]{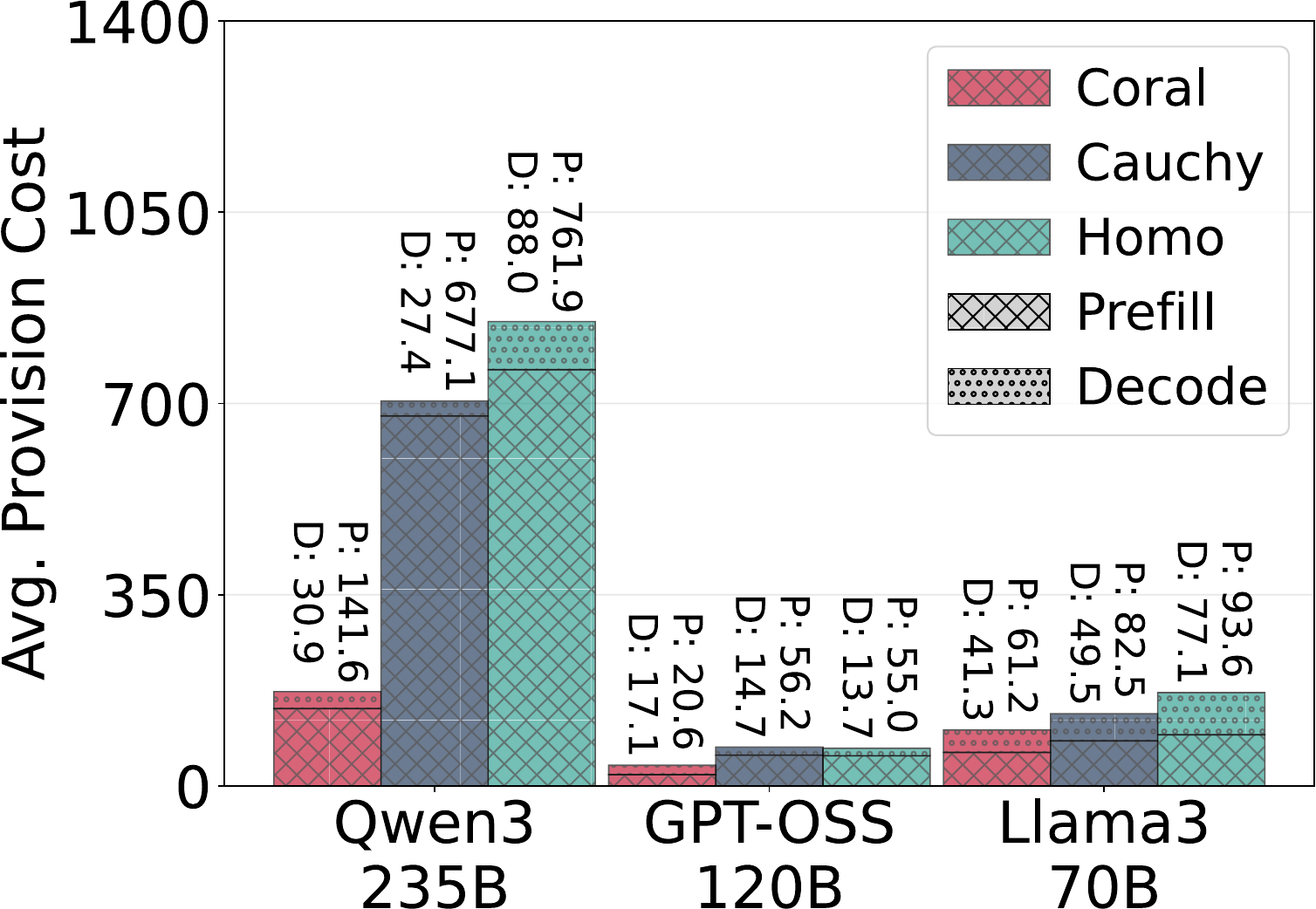}
        \caption{Per-model average provision cost breakdown in \largescale.
        }
        \label{fig:e2e_exp1_d}
    \end{subfigure}
    \caption{\textbf{Hourly cost comparison under default settings.} (a, c) Hourly cost per epoch across both deployment scales. (b, d) Per-model average provisioning cost decomposed into prefill (P) and decode (D) phases. The \largescale breakdown displays only the three largest models, which dominate total expenditure.}
    \label{fig:e2e_exp1}
    \Description{Hourly cost comparison under default settings across the two model and GPU setups.}
\end{figure*}
\begin{figure}
    \centering
    \begin{subfigure}[t]{0.48\linewidth}
        \centering
        \includegraphics[width=\linewidth]{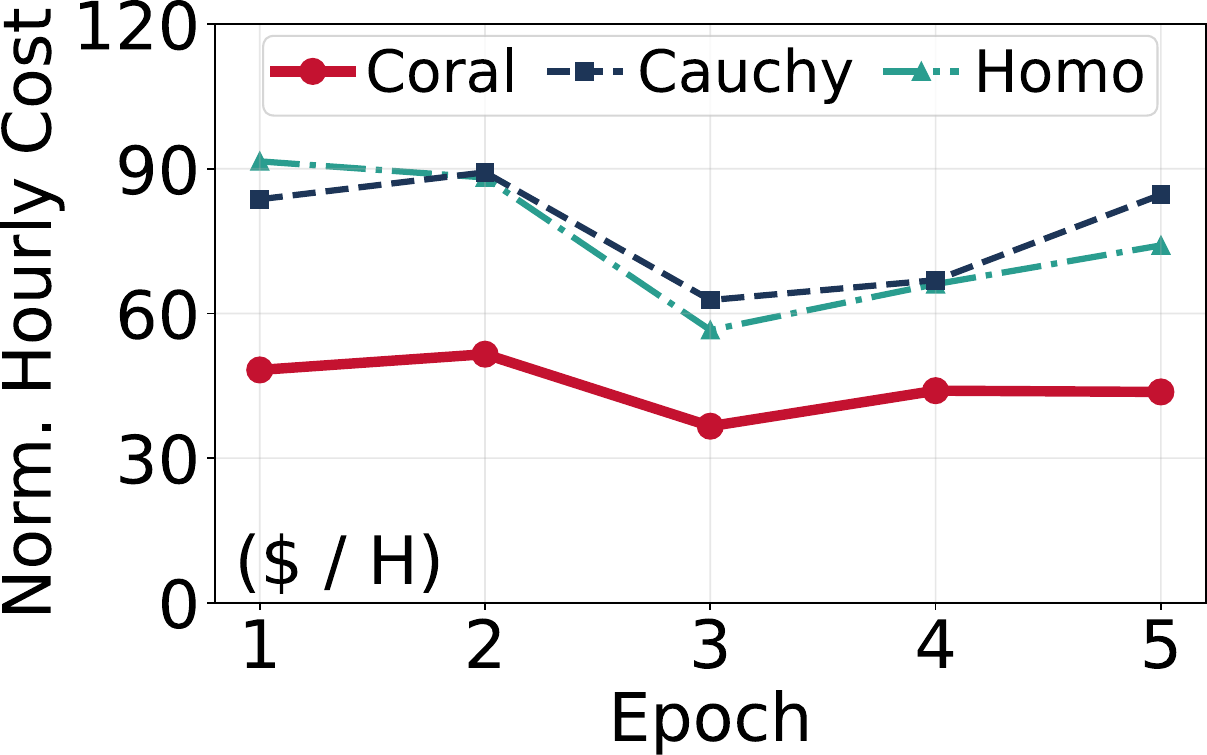}
        \caption{\Smallscale.}
        \label{fig:e2e_exp2_a}
    \end{subfigure}
    \hfill
    \begin{subfigure}[t]{0.48\linewidth}
        \centering
        \includegraphics[width=\linewidth]{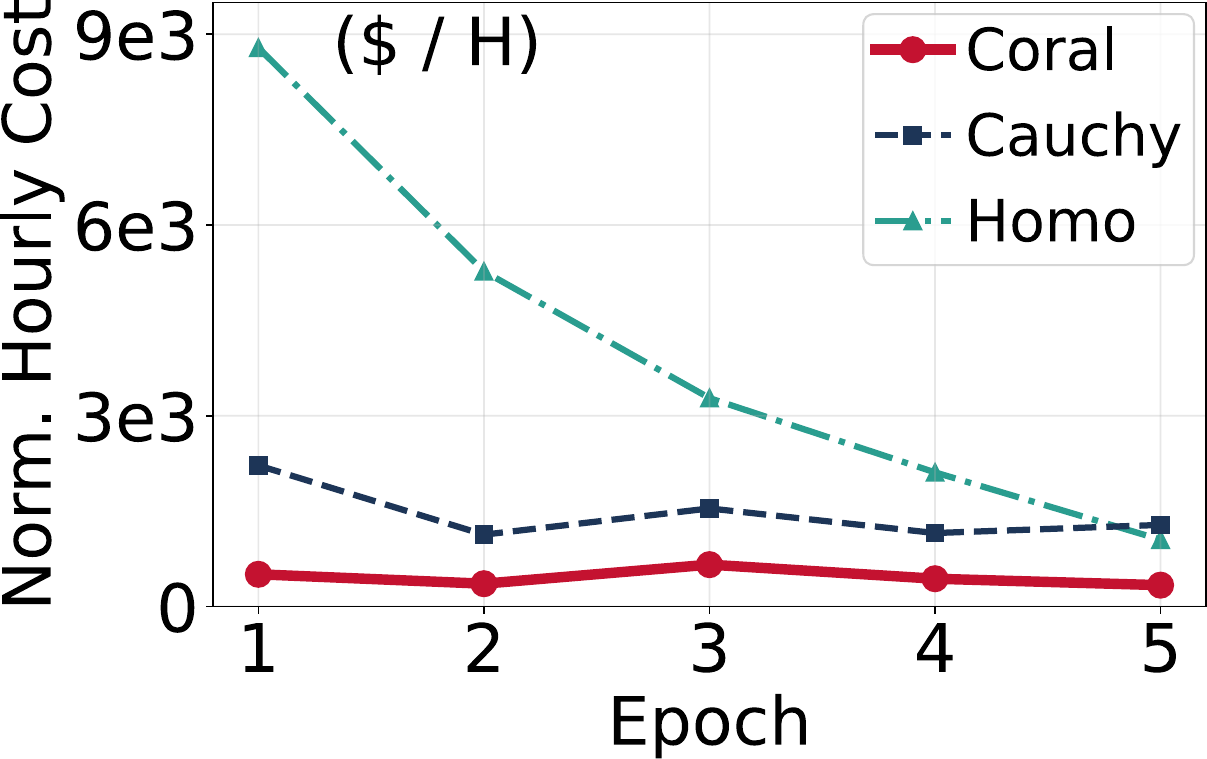}
        \caption{\Largescale.}
        \label{fig:e2e_exp2_b}
    \end{subfigure}
    \caption{
    Goodput-normalized hourly cost under tight resource availability.
    }
    \label{fig:e2e_exp2}
    \Description{Hourly cost under scarce resource availability.}
\end{figure}
\begin{figure}
    \centering
    \includegraphics[width=0.98\linewidth]{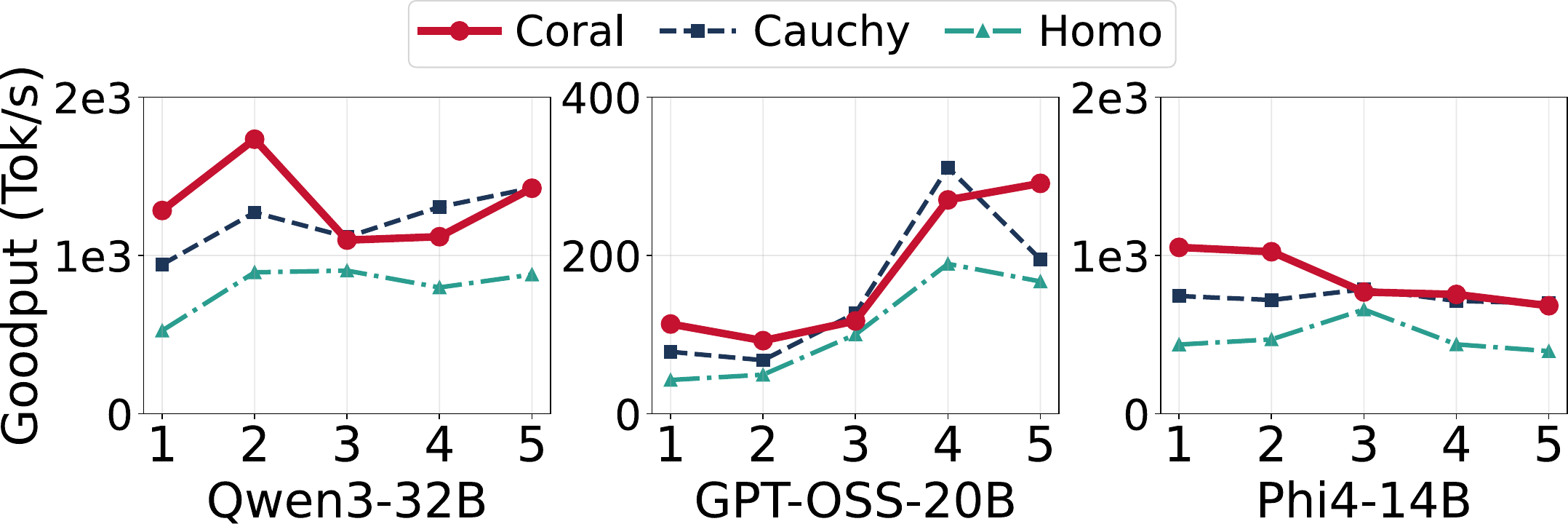}
    \caption{
    Decode goodput across epochs under tight resource availability (\smallscale). Prefill follows the same trend. 
    }
    \label{fig:e2e_exp2_goodput_core}
    \Description{Decode goodput across epochs under tight resource availability (\smallscale).}
\end{figure}

\myparagraph{Resource Setup.} GPU availability is derived from Alibaba's production cluster traces~\cite{duan2026GFS}. By default, we scale availability high enough to guarantee feasibility for all methods. For our resource-constrained study (Sec.~\ref{sec:e2e_exp2}), we throttle availability to a tight margin: 25\% (\smallscale) and 75\% (\largescale) above estimated minimum demand. Node pricing strictly mirrors public on-demand rates for AWS US-East-2, AWS AP-Northeast-2, and GCP US-Central-1.

\myparagraph{Evaluation Duration.} Experiments run for 30 minutes, dynamically reconfiguring the cluster every 6 minutes (one \textbf{\emph{epoch}}). We cap experiments at this duration to control physical deployment costs. Because production clusters adjust less frequently, we calculate hourly cost by amortizing initialization penalties over a standard 60-minute interval.

\myparagraph{\sys Setup.} The offline Serving Template Library is generated on a 384-core AWS c8i instance, capping templates at $N_{\max} = 6$ nodes and memory capacity at $\rho = 12{\times}$ model size (Sec.~\ref{sec:sensitivity} analyzes sensitivity to $N_{\max}$ and $\rho$). This one-time process completes in a few hours. The online resource allocator executes on a 32-core c8i instance.

\myparagraph{Baselines.} We evaluate against two end-to-end baselines implemented atop the \sys runtime for fairness. \textbf{\homo} restricts each replica to homogeneous hardware, though hardware can vary across replicas (the same assumption used in SkyServe~\cite{mao2025skyserve} and SageServe~\cite{jaiswal2025sageserve}). It greedily provisions the most cost-efficient (highest goodput per USD) homogeneous strategy per model. \textbf{Cauchy} is adapted from the PD-disaggregated serving system of the same name~\cite{zhang2025cauchy}. We retain its cost-efficiency model and resource-allocation algorithm, and extend its GPU-combo definition to allow a single prefill replica to feed multiple decode replicas, granting it more flexibility for multi-LLM serving. Heterogeneity-aware placement systems like \textbf{Helix}~\cite{mei2025helix} do not perform resource allocation, so we compare against Helix separately in Sec.~\ref{sec:helix_comparison}.

\myparagraph{Evaluation Metrics.} Our primary metric is \emph{hourly cost} (USD/h), which aggregates machine provisioning and amortized initialization costs. For contention studies (Sec.~\ref{sec:e2e_exp2}), we additionally report \emph{goodput}: the rate of generated tokens (Tok/s) that successfully meet their latency SLO.


\begin{figure*}
    \centering
    \includegraphics[width=0.96\linewidth]{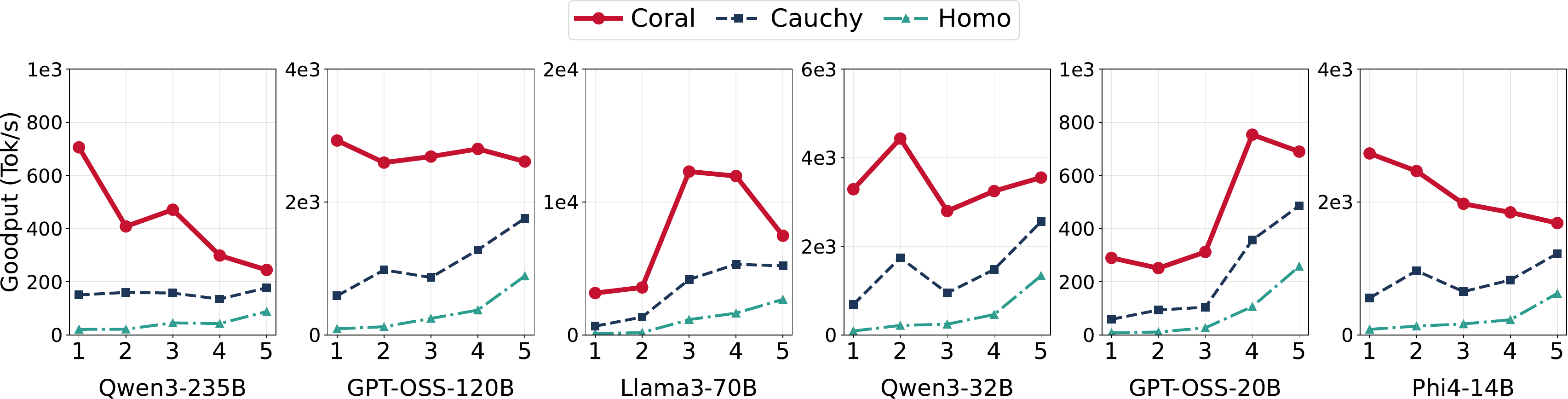}
    \caption{
    Decode goodput across epochs under tight resource availability (\largescale). Prefill follows the same trend.
    }
    \label{fig:e2e_exp2_goodput_extended}
    \Description{Decode goodput across epochs under tight resource availability (\largescale).}
\end{figure*}

\subsection{Simulator Fidelity}
\label{sec:simulator_fidelity}

To validate the simulator, we replicate all real-hardware experiments from Sec.~\ref{sec:e2e_exp1} and Sec.~\ref{sec:e2e_exp2}. We use per-request latency as our primary fidelity metric: because aggregate metrics like goodput and SLO attainment derive directly from latency, matching the underlying latency distribution guarantees accuracy across any SLO threshold. Across all setups, simulated prefill and decode latencies deviate from real hardware by an average of only 5.6\% and 7.2\%, respectively. Fig.~\ref{fig:simulator_fidelity} illustrates a representative comparison for Phi-4 14B, demonstrating tight alignment across the full latency CDF, including the tail. We further validate the most cost-efficient Serving Templates for all three large models introduced in the \largescale (Qwen-3 235B, GPT-OSS 120B, Llama-3 70B), observing an average latency deviation of under 2\%. Together, these results confirm the simulator's high fidelity as a proxy for our large-scale evaluations.


\subsection{End-to-End Cost Efficiency}
\label{sec:e2e_exp1}

This section evaluates whether \sys reduces total serving costs compared to state-of-the-art baselines across diverse model mixtures and heterogeneous GPU pools (Fig.~\ref{fig:e2e_exp1}).

In the \smallscale (Fig.~\ref{fig:e2e_exp1_a}), \sys reduces average hourly costs by $1.62\times$ over \homo and $1.60\times$ over Cauchy, with the online ILP resolving in 0.24s on average. As the per-model breakdown in Fig.~\ref{fig:e2e_exp1_b} shows, Qwen-3 32B drives roughly 60\% of the total expenditure because it is the largest model and lacks the compute-saving MoE or hybrid attention mechanisms. \sys yields its highest savings on this model ($2.02\times$ over \homo, $1.88\times$ over Cauchy), primarily during the prefill phase. Examining the selected prefill configurations, both baselines restrict replicas to single nodes (e.g., one replica per $2{\times}$L40S or $4{\times}$L4 node), varying the node mix across epochs to track demand. Conversely, \sys deploys heterogeneous templates that fuse L4 and L40S nodes into a single replica (e.g., one $1{\times}$L4 plus three $1{\times}$L40S nodes), employing non-uniform pipeline partitioning and intra-stage data parallelism.

In the \largescale (Fig.~\ref{fig:e2e_exp1_c}), \sys achieves even greater savings, reducing average hourly costs by $3.14\times$ over \homo and $2.66\times$ over Cauchy, with an average ILP solve time of 9.68 seconds. Qwen-3 235B again dominates costs (Fig.~\ref{fig:e2e_exp1_d}) due to its massive parameter count, despite its MoE architecture. \sys reduces Qwen-3 235B serving cost by $4.93\times$ over \homo and $4.08\times$ over Cauchy, again driven by prefill optimizations. A similar architectural pattern emerges: while baselines confine replicas to monolithic $8{\times}$A100 or $8{\times}$H100 nodes, \sys instead uses templates that combine multiple smaller 1- and 2-GPU A100 and H100 nodes into one replica, optimizing capacity through non-uniform pipelining.

These results confirm that intra-replica heterogeneity is critical for cost efficiency; mixing GPU types allows \sys to tailor instances to specific latency SLOs and throughput demands (Sec.~\ref{sec:opportunities}). Finally, across all methods and scales, amortized initialization overhead remains below 1\% of total hourly cost, showing that the allocator's initialization penalty successfully suppresses unnecessary cluster churn.


\subsection{Goodput Under Tight Resource Availability}
\label{sec:e2e_exp2}

\begin{figure*}
    \centering
    \begin{subfigure}[t]{0.24\linewidth}
        \centering
        \includegraphics[width=\linewidth]{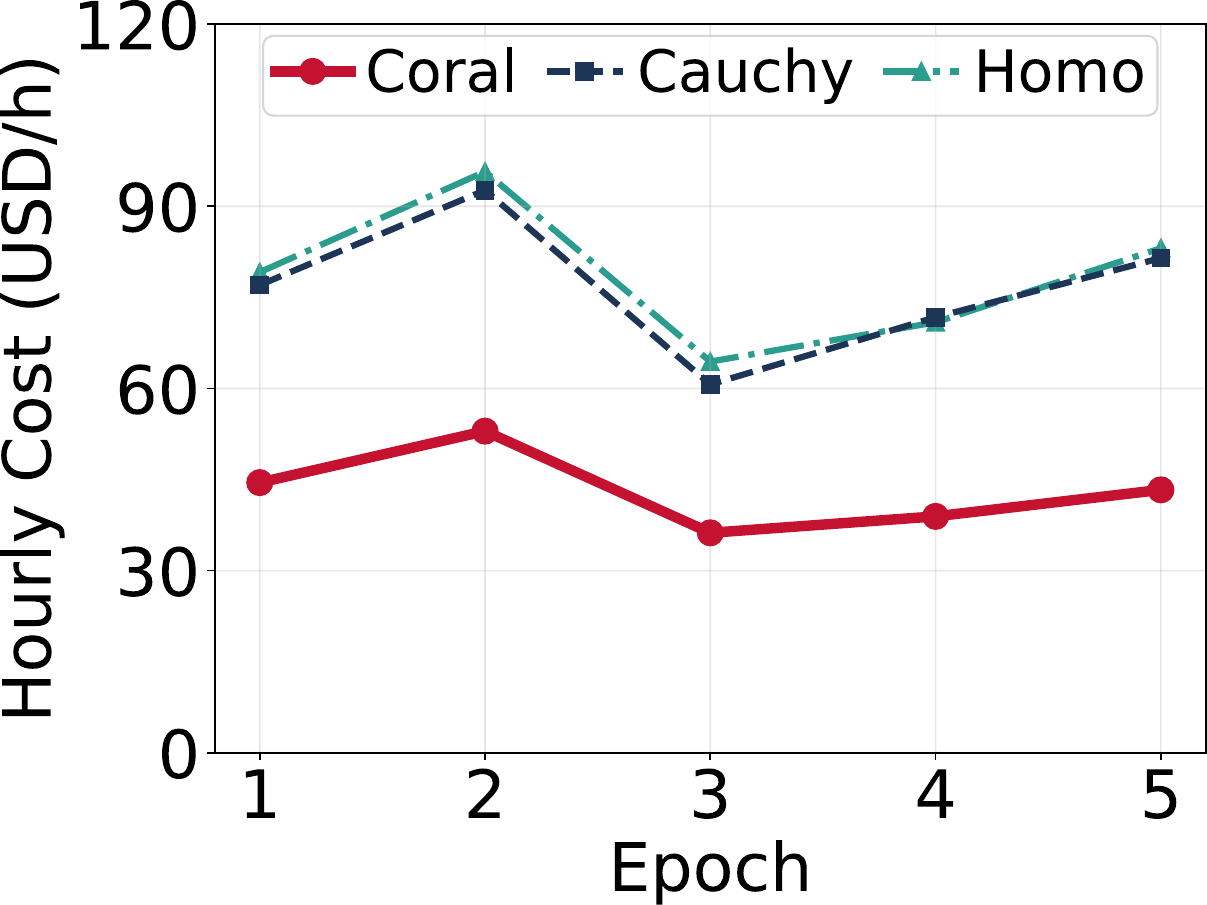}
        \caption{Large-Heavy, \smallscale.}
        \label{fig:e2e_exp3_a}
    \end{subfigure}
    \hfill
    \begin{subfigure}[t]{0.24\linewidth}
        \centering
        \includegraphics[width=\linewidth]{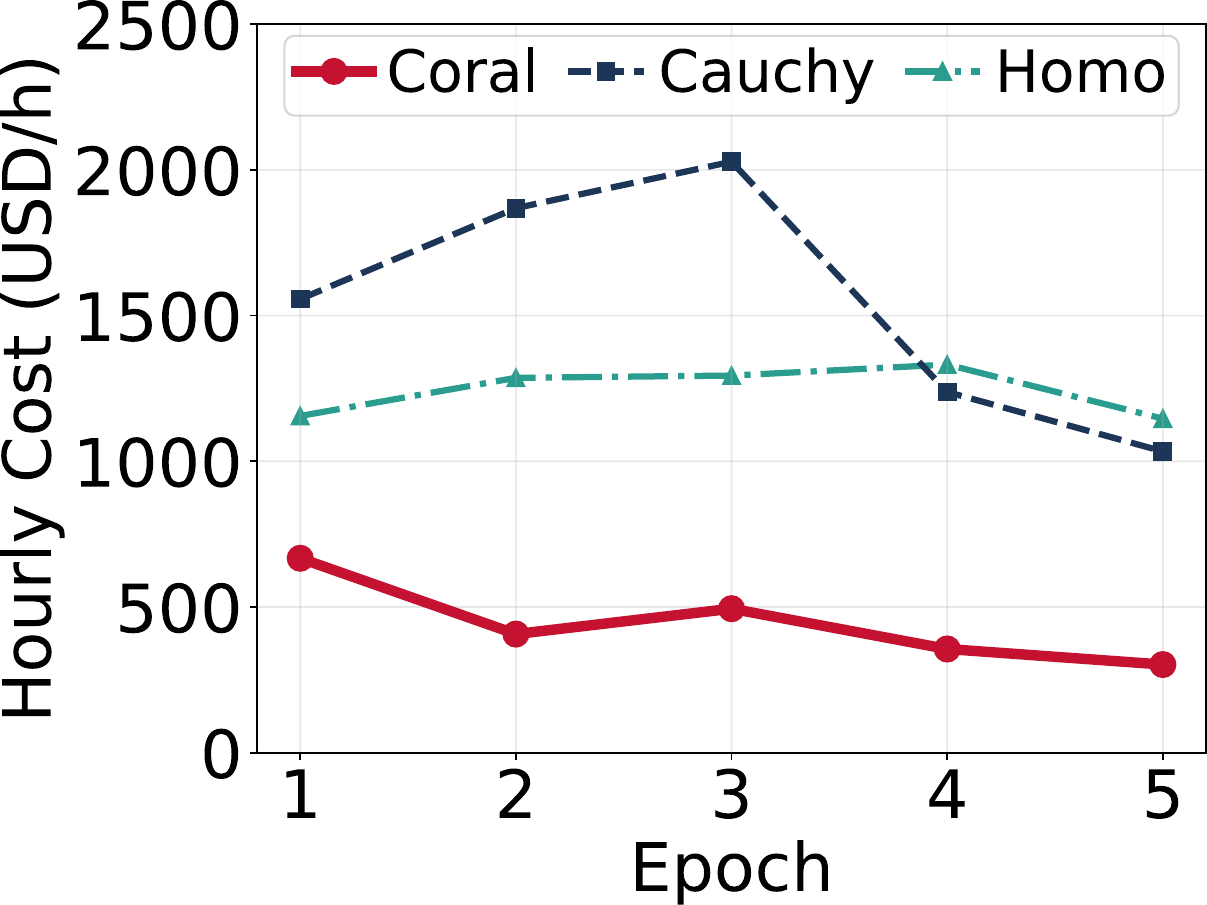}
        \caption{Large-Heavy, \largescale.}
        \label{fig:e2e_exp3_b}
    \end{subfigure}
    \hfill
    \begin{subfigure}[t]{0.24\linewidth}
        \centering
        \includegraphics[width=\linewidth]{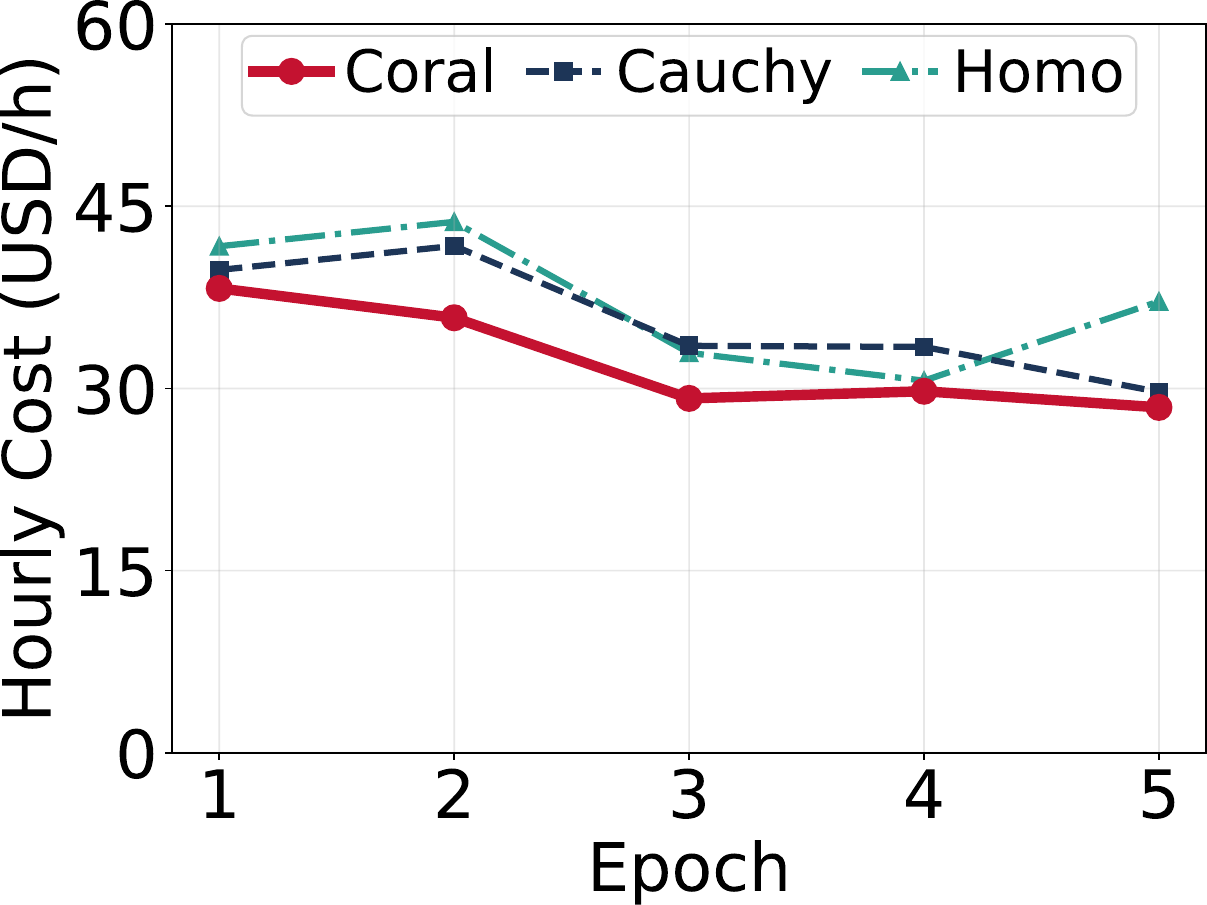}
        \caption{Small-Heavy, \smallscale.}
        \label{fig:e2e_exp3_c}
    \end{subfigure}
    \hfill
    \begin{subfigure}[t]{0.24\linewidth}
        \centering
        \includegraphics[width=\linewidth]{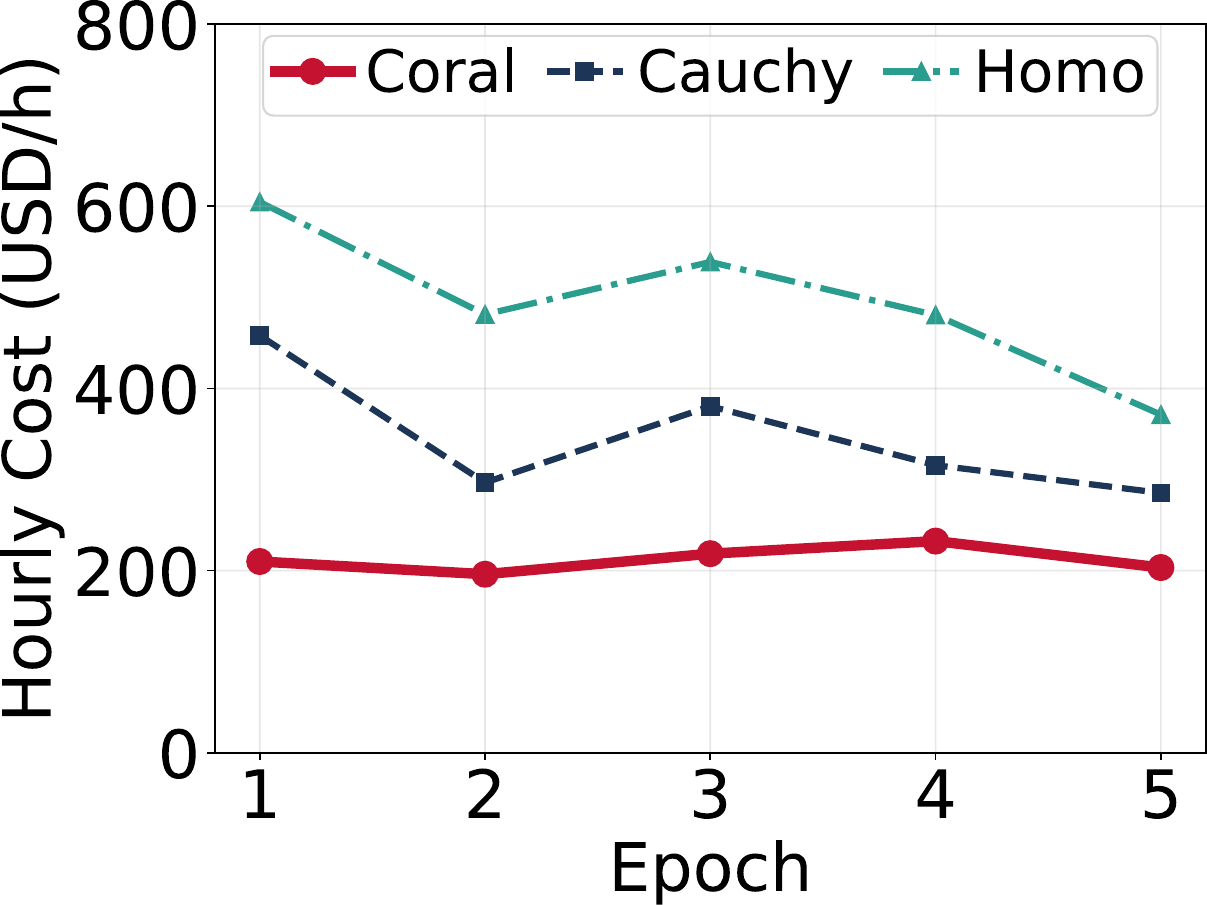}
        \caption{Small-Heavy, \largescale.}
        \label{fig:e2e_exp3_d}
    \end{subfigure}
    \caption{\textbf{Hourly cost under imbalanced demand.} 80\% of requests are routed to either the largest (Large-Heavy) or smallest (Small-Heavy) third of models. \sys's cost advantage expands significantly when large models dominate the workload.}
    \label{fig:e2e_exp3}
    \Description{Hourly cost under imbalanced demand.}
\end{figure*}

This section evaluates whether \sys sustains high goodput and mitigates contention under tight availability. If a method cannot satisfy all models' throughput demands, we uniformly throttle its per-model arrival rate until a feasible solution exists, preserving the balanced-demand assumption.

In the \smallscale (Fig.~\ref{fig:e2e_exp2_a}, \ref{fig:e2e_exp2_goodput_core}), \sys delivers 1.66$\times$ higher average goodput than \homo, reducing the goodput-normalized hourly cost by 1.67$\times$. Against Cauchy, \sys achieves 1.12$\times$ higher goodput and a 1.73$\times$ reduction in normalized cost. The online ILP resolves in just 0.11 seconds on average.

Compared to its solution in Sec.~\ref{sec:e2e_exp1}, \sys's cluster setup is 10\% more expensive on average. During the tightest interval (epoch~1), \sys redirects three high-end L40S GPUs from Phi-4 to Qwen-3---the heavier model that critically requires them---while shifting Phi-4 to A10G and L4 nodes. To achieve this, \sys expands its active heterogeneous templates from 1 to 4. Conversely, the baselines lack cross-model coordination: \homo optimizes each model in isolation, and Cauchy rigidly encodes per-model efficiency in its objective. Consequently, both baselines force all models to contend for the scarce L40S GPUs, leaving none with sufficient capacity.

In the \largescale (Fig.~\ref{fig:e2e_exp2_b}, \ref{fig:e2e_exp2_goodput_extended}), \sys drastically outperforms the baselines. Against \homo, \sys delivers an overwhelming 7.64$\times$ higher average goodput while reducing goodput-normalized hourly cost by 8.99$\times$. Against Cauchy, it achieves 2.39$\times$ higher goodput and a 3.27$\times$ reduction in normalized cost. The online ILP resolves in 41s on average.

Here, \sys's allocation is 25\% more expensive on average than in Sec.~\ref{sec:e2e_exp1}. The tightest interval (epoch~3) also requires the longest solve time (2 minutes) and incurs the largest cost increase (53\%). During this epoch, \sys scales its heterogeneous instances from 9 to 19. To sustain throughput for the three largest models (Qwen-3 235B, GPT-OSS 120B, Llama-3 70B), \sys deploys 16 distinct heterogeneous instances---up from 6 in Sec.~\ref{sec:e2e_exp1}, where only Qwen-3 235B relied heavily on them. This intricate cross-model coordination drives the longer solve time. As in the \smallscale, the baselines fail because models greedily compete for scarce A100 and H100 nodes, resulting in severe resource starvation.

These results yield two key takeaways: first, global joint optimization is indispensable for satisfying aggregate demand under severe resource scarcity; second, heterogeneous Serving Templates provide the solver with the critical flexibility needed to resolve cross-model hardware contention.


\begin{figure}
    \centering
    \includegraphics[width=0.98\linewidth]{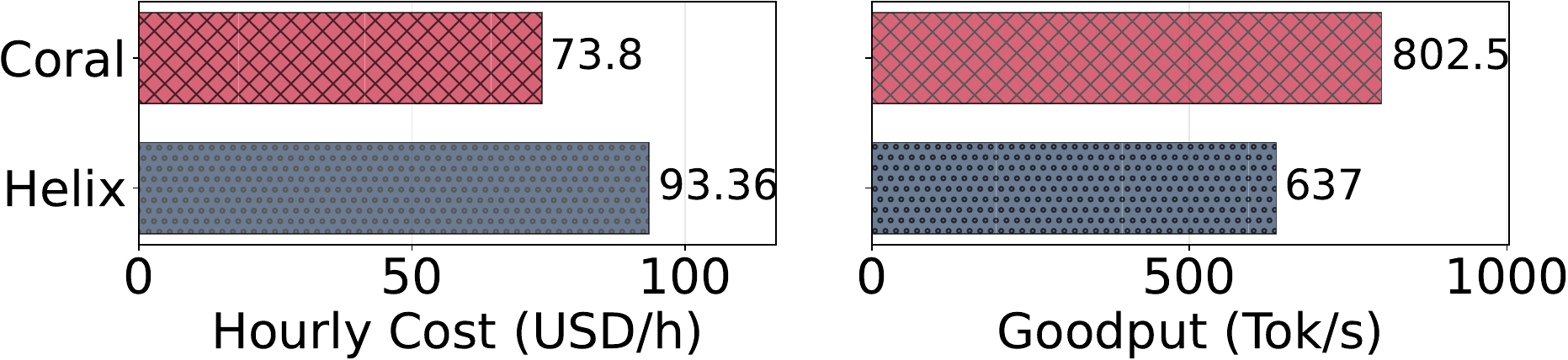}
    \caption{
    Comparison with Helix on Helix's "High GPU-Heterogeneity Cluster" setup.
    }
    \label{fig:helix_comparison}
    \Description{Comparison with Helix.}
\end{figure}

\subsection{Robustness to Demand Imbalance}
\label{sec:e2e_exp3}

This section evaluates \sys's resilience to skewed throughput demands across models. We consider two imbalanced distributions: \emph{Large-Heavy} routes 80\% of requests to the top third of models by size (Qwen-3 32B in \smallscale; Qwen-3 235B and GPT-OSS 120B in \largescale), while \emph{Small-Heavy} routes 80\% to the bottom third (Phi-4 14B in \smallscale; Phi-4 14B and GPT-OSS 20B in \largescale).

\myparagraph{Large-Heavy (Fig.~\ref{fig:e2e_exp3_a}, \ref{fig:e2e_exp3_b}).} 
In the \largescale, \sys reduces average hourly costs by $2.79\times$ over \homo and $3.47\times$ over Cauchy; in the \smallscale, it achieves reductions of $1.82\times$ and $1.78\times$, respectively. In this regime, the large models dominate spending, consuming $\sim$80\% of the total budget. Because these massive models require multi-GPU replicas, \sys has ample opportunity to exploit intra-replica heterogeneity, driving large cost savings.

\myparagraph{Small-Heavy (Fig.~\ref{fig:e2e_exp3_c}, \ref{fig:e2e_exp3_d}).} 
In the \largescale, \sys lowers hourly costs by $2.34\times$ over \homo and $1.64\times$ over Cauchy. Despite handling only a fraction of the traffic, Qwen-3 235B and Llama-3 70B are so large that they still account for 40\% and 20\% of the total cost, respectively, allowing \sys to extract substantial savings from them. However, in the \smallscale, all three methods yield nearly identical costs, with \sys only $\sim$10\% cheaper on average. This is expected: the cost-dominating Phi-4 14B fits on 1--2 GPUs, leaving minimal room for \sys to perform intra-replica optimization.


\subsection{Single-Model Comparison with Helix}
\label{sec:helix_comparison}

This section compares \sys against Helix~\cite{mei2025helix}, which optimizes model placement for a single model on a \textbf{fixed} heterogeneous cluster via MILP.

We adopt Helix's largest experimental setup (the ``High GPU-Heterogeneity Cluster'') and bias the configuration heavily in Helix's favor. \sys allocates from the exact same cluster Helix uses (4$\times$A100 40G, 6$\times$V100 16G, 16$\times$L4 24G, and 38$\times$T4 16G) to serve the 70B Llama model under AWS US-East-2 pricing. We drive the workload at 4~req/s---exceeding Helix's reported maximum throughput---and constrain \sys with prefill and decode SLOs of 2090~ms and 730~ms, matching the median latencies achieved by Helix. Helix itself runs without SLO constraints. To avoid prohibitive hardware costs, we evaluate \sys using our high-fidelity simulator and compare directly against Helix's reported simulated results.

As Fig.~\ref{fig:helix_comparison} demonstrates, \sys reduces costs by 21\% and improves throughput by 26\% while strictly satisfying both SLOs. \sys achieves this by dedicating the A100 nodes to prefill and partitioning the remaining hardware into three independent decode instances (combining L4 and T4 nodes), intentionally idling six V100s and one T4 to minimize cost. Conversely, Helix forces all 64 GPUs into a single monolithic pipeline, incurring severe cross-stage communication bottlenecks. By decomposing the resource pool into smaller, right-sized Serving Instances, \sys eliminates the overhead of massive heterogeneous pipelines, delivering higher performance at a lower cost.


\subsection{Ablation Study}
\label{sec:ablation}

\begin{figure}
    \centering
    \includegraphics[width=\linewidth]{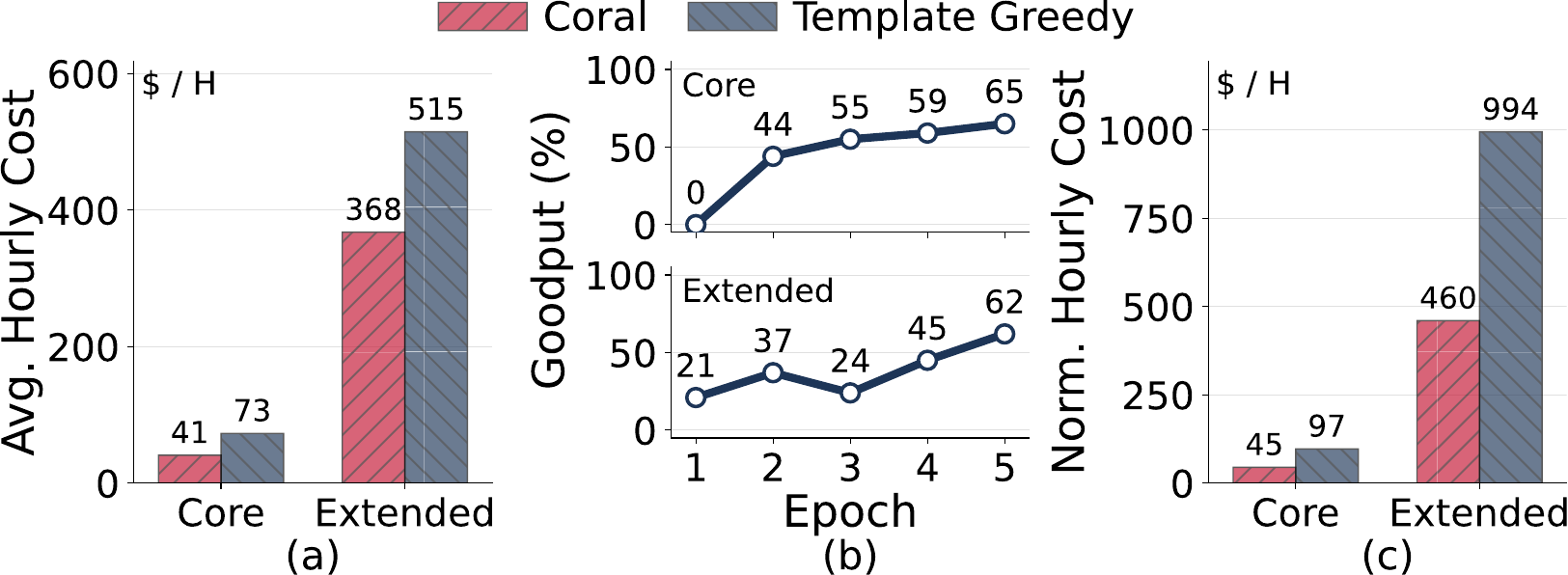}
    \caption{\textbf{Impact of global allocation.} \textbf{(a)} Avg. hourly cost under \textit{abundant availability}. \textbf{(b, c)} Goodput and goodput-normalized hourly cost under \textit{restricted availability}.}
    \label{fig:ablation_greedy}
    \Description{Ablation study with Template Greedy.}
\end{figure}

To isolate the impact of the online resource allocation ILP (Sec.~\ref{sec:resource_allocation}), we ablate \sys against a \textit{Template Greedy} baseline that independently selects the most cost-efficient Serving Template per model. We evaluate both core and extended setups under abundant (Sec.~\ref{sec:e2e_exp1}) and restricted (Sec.~\ref{sec:e2e_exp2}) hardware availability. Under abundant availability, \sys's global optimization reduces average hourly costs by $1.78\times$ in the \smallscale and $1.40\times$ in the \largescale compared to greedy selection (Fig.~\ref{fig:ablation_greedy}(a)). Under restricted availability, the greedy approach severely degrades, achieving only 45\% and 38\% of \sys's goodput on average across the two setups, and completely failing to find a feasible solution during one \smallscale epoch (Fig.~\ref{fig:ablation_greedy}(b)). \sys reduces the goodput-normalized hourly cost by $2.16\times$ across both setups (Fig.~\ref{fig:ablation_greedy}(c)). These results confirm that global, cross-model coordination is strictly necessary to navigate hardware contention and maximize cost efficiency.


\subsection{Sensitivity of Template Pruning}
\label{sec:sensitivity}

\begin{figure}
    \centering
    \includegraphics[width=\linewidth]{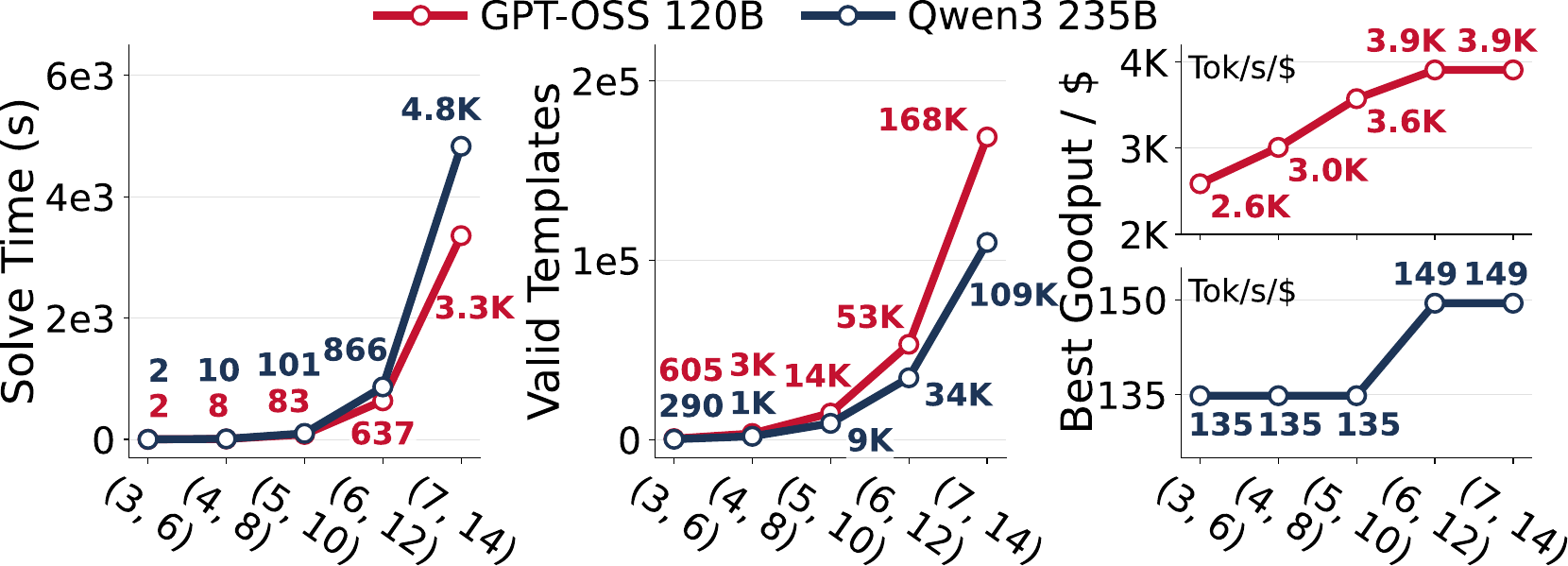}
    \caption{\textbf{Template pruning sensitivity.} Although solve time and template count scale exponentially with pruning parameters $(N_{\max}, \rho)$, best cost efficiency plateaus at $(6, 12)$.}
    \label{fig:sensitivity_analysis}
    \Description{Sensitivity of Serving Template generation.}
\end{figure}

This section evaluates the sensitivity of the Serving Template generator to its two pruning parameters (Sec.~\ref{sec:template_generation}): the node cap $N_{\max}$ and the memory capacity ratio $\rho$. Because adding nodes inherently inflates aggregate memory, we sweep both parameters jointly. Using the GPT-OSS 120B prefill and Qwen-3 235B decode phases as a testbed, we execute the generator on a 384-core AWS c8i instance. For each $(N_{\max}, \rho)$ configuration, we measure the total valid templates produced, the aggregate solve time, and the best cost efficiency (Goodput/USD) achieved.

As Fig.~\ref{fig:sensitivity_analysis} demonstrates, while both template count and solve time scale exponentially with $(N_{\max}, \rho)$, maximum cost efficiency plateaus at $(N_{\max}, \rho) = (6, 12)$. This is expected: large node sets suffer from higher inter-node communication overhead and are dominated by partitioning those same resources into multiple smaller, independent replicas. Searching beyond this plateau needlessly inflates offline compute costs without discovering more efficient templates. These results empirically validate the principled pruning strategy proposed in Sec.~\ref{sec:template_generation} and justify our default configuration of $(N_{\max}, \rho) = (6, 12)$ for all end-to-end evaluations.


\section{Related Work}
\label{sec:related-work}

\myparagraph{Prefill-Decode (PD) Disaggregated LLM Serving.}
PD disaggregation provisions prefill and decode independently to exploit their distinct execution characteristics~\cite{zhong2024distserve,patel2024splitwise,zhang2025cauchy,hu2024inference,qin2024mooncake}. Splitwise~\cite{patel2024splitwise} and DistServe~\cite{zhong2024distserve} pioneer this separation. Cauchy~\cite{zhang2025cauchy} extends it by assigning different GPUs to the prefill and decode pools. However, these systems confine heterogeneity strictly to the phase boundary; individual replicas remain entirely homogeneous. \sys fundamentally extends this paradigm by enabling \emph{intra-replica} heterogeneity via hybrid pipeline and data parallelism, seamlessly supporting both PD-disaggregated and aggregated serving (Sec.~\ref{sec:system_design}).

\myparagraph{Resource Allocation for LLM Serving.}
Another line of work scales serving resources dynamically against fluctuating demand, pricing, and availability~\cite{miao2024spotserve,mao2025skyserve,jaiswal2025sageserve}. SpotServe~\cite{miao2024spotserve} leverages preemptible instances, SkyServe~\cite{mao2025skyserve} autoscales replicas across multi-cloud regions, and SageServe~\cite{jaiswal2025sageserve} integrates workload forecasting to minimize reconfiguration overhead. Crucially, while these systems optimize cluster-level provisioning and routing, they treat each replica as a rigid, homogeneous block. \sys operates orthogonally: by jointly optimizing hardware selection and model placement, it unlocks the cost efficiency of intra-replica heterogeneity that prior allocators leave on the table.

\myparagraph{Heterogeneous LLM Serving.}
Prior work leverages GPU heterogeneity across multiple dimensions. M\'elange~\cite{griggs2024m} shows that optimal GPU choice depends on workload and SLOs, but restricts replicas to a single homogeneous type. BOute~\cite{jiang2026boute} routes requests across fixed heterogeneous hardware to trade off latency and quality. Others optimize composition and placement but minimize offline batch makespan on static pools~\cite{jiang2025demystifying}—the dual of \sys's problem, without online reconfiguration. Systems like Helix~\cite{mei2025helix} and HexGen~\cite{jiang2023hexgen} optimize placement on \emph{fixed} heterogeneous clusters, abstracting away resource provisioning. Similarly, OpWeave~\cite{li2026opweave} jointly optimizes operator partitioning and deployment configurations across heterogeneous GPUs, but assumes a given hardware pool. These approaches do not address dynamic resource provisioning (and as Sec.~\ref{sec:solution_overview} shows, wrapping placement optimization in an online allocation loop is intractable). \sys is the first to jointly provision dynamic heterogeneous resources \emph{and} orchestrate model placement across them for multiple LLMs under strict latency SLOs.

\myparagraph{Multi-Model LLM Serving.}
Recent systems also target the concurrent execution of multiple LLMs. Prism~\cite{yu2025prism} dynamically redistributes memory among colocated models, Aegaeon~\cite{xiang2025aegaeon} applies token-granularity autoscaling, and FlexPipe~\cite{lin2025flexpipe} multiplexes models via dynamic pipeline refactoring. While these systems excel at sharing and scheduling within a \emph{fixed} hardware pool, they assume the underlying cluster composition is pre-determined. \sys solves the upstream challenge: determining exactly which heterogeneous resources to provision and how to place models across them. Consequently, these fine-grained runtime mechanisms could be naturally layered atop \sys's Serving Instances.

\section{Conclusion}
\label{sec:conclusion}

This paper presents \sys, an adaptive, heterogeneity-aware system for cost-efficient multi-LLM serving. \sys jointly optimizes resource allocation and model placement via a lossless two-stage decomposition: offline ILP-based \emph{Serving Template} generation and online template instantiation. Removing the heavy placement search from the critical path accelerates online scheduling from hours to tens of seconds, enabling the cluster to dynamically adapt to fluctuating demands and hardware availability. Evaluated across 6 models and 20 GPU configurations, \sys reduces serving costs by up to 2.79$\times$ and delivers up to 2.39$\times$ higher goodput under severe resource constraints.

\section*{Acknowledgments}
Generative AI tools, including Codex, Claude Code, and Gemini, were utilized to assist with code generation during system implementation and to refine the prose of this manuscript. This research was supported in part by NSF awards CNS-2211882 and CNS-2239351, Sloan Foundation faculty fellowships, and research awards from Amazon, Cisco, Google, Jane Street, Meta, NVIDIA, Oracle, Qualcomm, and Samsung. We also thank the members and companies of the PDL consortium (Bloomberg LP, Everpure, Google, Jane Street, LayerZero Labs, Meta, Microsoft Research, Oracle Corporation, Salesforce, Uber, Western Digital) for their interest, insights, feedback, and support. We thank Lambda for providing cloud computing credits that supported the experiments conducted in this work.

\bibliographystyle{ACM-Reference-Format}
\bibliography{ref}

\end{document}